\documentclass[letterpaper,twocolumn,10pt]{article}
\usepackage{usenix}

\usepackage{tikz}
\usepackage{amsmath}

\usepackage{filecontents}
\usepackage[utf8]{inputenc}
\usepackage[T1]{fontenc}
\usepackage{hyperref}
\usepackage{url}
\usepackage{booktabs}
\usepackage{amsfonts}
\usepackage{nicefrac}
\usepackage{microtype}
\usepackage{xcolor}
\usepackage{amsfonts}
\usepackage{algorithm}
\usepackage{algpseudocode}
\usepackage{caption}
\usepackage{subcaption}
\usepackage{graphicx}
\usepackage{xspace}
\usepackage{enumitem}
\usepackage{multirow}
\usepackage{makecell}
\usepackage[table]{xcolor}
\usepackage{array}
\usepackage{listings}
\usepackage{float}
\usepackage{placeins}
\usepackage{tabularx}
\usepackage{makecell}
\usepackage{longtable}
\usepackage{pifont}
\newcommand{\agent}[1]{\ding{\numexpr171+#1\relax}}
\newcommand{\tool}{\textit{AutoEG}\xspace}

\begin{document}

\date{}

\title{\Large \bf \tool: Exploiting Known Third-Party Vulnerabilities in \\Black-Box Web Applications}

\newcommand{\mksmu}[0]{{{$^1$}}}
\newcommand{\mkntu}[0]{{{$^2$}}}
\newcommand{\mktju}[0]{{{$^3$}}}

\author{
    {\rm Ruozhao Yang}\mksmu \rm ,
    {\rm Mingfei Cheng}\mksmu \rm ,
    {\rm Gelei Deng}\mkntu \rm ,
    {\rm Junjie Wang}\mktju \rm ,
    {\rm Tianwei Zhang}\mkntu \rm ,
    {\rm Xiaofei Xie}\mksmu \rm \\ 
    \mksmu {Singapore Management University},
    \mkntu {Nanyang Technological University},
    \mktju {Tianjin University}
    \\
    \medskip
}

\maketitle

\begin{abstract}
Large-scale web applications are widely deployed with complex third-party components, inheriting security risks arising from component vulnerabilities. Security assessment is therefore required to determine whether such known vulnerabilities remain practically exploitable in real applications. Penetration testing is a widely adopted approach that validates exploitability by launching concrete attacks against known vulnerabilities in real-world black-box systems. However, existing approaches often fail to automatically generate reliable exploits, limiting their effectiveness in practical security assessment. This limitation mainly stems from two issues: (1) precisely triggering vulnerabilities with correct technical details, and (2) adapting exploits to diverse real-world deployment settings.


In this paper, we propose \tool, a fully automated multi-agent framework for exploit generation targeting black-box web applications. \tool has two phases: First, \tool extracts precise vulnerability trigger logic from unstructured vulnerability information and encapsulates it into reusable trigger functions. Second, \tool uses trigger functions for concrete attack objectives and iteratively refines exploits through feedback-driven interaction with the target application. We evaluate \tool on 104 real-world vulnerabilities with 29 attack objectives, resulting in 660 exploitation tasks and 55,440 exploit attempts. \tool achieves an average success rate of 82.41\%, substantially outperforming state-of-the-art baselines, whose best performance reaches only 32.88\%.


\end{abstract}

\section{Introduction}

Web applications have become a fundamental infrastructure of modern online services, supporting critical domains such as banking, e-commerce, and social networking, and consequently attracting sustained attention from attackers~\cite{li2024fuzzcache}. Modern web applications are typically built upon multiple interacting components, such as web frameworks, middleware, and databases~\cite{lauinger2018thou}. This dependency complexity significantly increases security risks. Individual components may contain vulnerabilities, and the rapid evolution of third-party components often outpaces application-level maintenance, leaving deployed applications exposed to known vulnerabilities. Prior studies show that 81.5\% of software projects rely on outdated third-party components~\cite{jia2025impact}. Because the technical details of known vulnerabilities are publicly disclosed, these outdated dependencies present readily exploitable attack surfaces, often leading to severe security consequences. This reality highlights the necessity of systematically evaluating whether known vulnerabilities remain triggerable and exploitable in real-world web applications, namely known-vulnerability-based web application penetration testing.

Penetration testing evaluates the security of deployed applications by emulating real-world attacks under authorized and black-box settings~\cite{scarfone2008technical,PTES}. 
The National Institute of Standards and Technology (NIST) characterizes the \textit{Attack} stage as ``the heart of any penetration test'', in which previously identified vulnerabilities are validated through concrete exploitation attempts against the target system~\cite{scarfone2008technical}. Accordingly, exploit generation constitutes a core task in known-vulnerability-based web application penetration testing.


In practice, generating working exploits for known vulnerabilities is non-trivial and involves two key requirements. 
(1) Correctly interpreting vulnerability trigger logic from vulnerability information and translating such information into concrete exploit implementations. The vulnerability information is obtained from public repositories (e.g., CVE and NVD), whose records typically include free-form external references and are therefore fragmented and unstructured, making them difficult to work with.
(2) Precisely adapting the generated exploit to the concrete deployment environment. In real-world deployments, environment-specific details such as URL paths, parameter names, request formats, and server-side behaviors vary across installations, and even minor deviations in these details can invalidate an exploit and cause execution failure.
This process is further complicated by the need to satisfy different attack objectives (i.e., high-level security goals that specify the intended impact of an exploitation attempt)~\cite{PTES}, which may impose varying exploitation requirements even for the same vulnerability. 
Therefore, exploit generation often relies on human expertise.


Existing approaches to web application penetration testing can be broadly categorized into manual testing and LLM-based automated testing. Manual penetration testing relies entirely on human experts to perform exploit generation through iterative crafting and refinement, resulting in limited efficiency and scalability~\cite{mu2018understanding}. Semi-automated approaches~\cite{pentestgpt} delegate exploit generation to LLMs but still require substantial human intervention during execution and adaptation. Fully automated approaches~\cite{shen2025pentestagent,kong2025vulnbotautonomouspenetrationtesting} employ dedicated LLM agents responsible for automated exploit generation; however, their success rates in real-world scenarios are reported below 40\%~\cite{jin2025good,yang2025pentesteval}. From the perspective of exploit generation, this limitation primarily stems from LLM-generated outputs, which frequently contain incorrect technical details for vulnerability triggering (e.g., missing critical parameters) and erroneous environment adaptation in real-world deployments (e.g., referencing non-existent paths or files).

In summary, while large language models have demonstrated strong capabilities in code generation and vulnerability analysis~\cite{nunez2024autosafecoder,fu2023chatgpt,nguyen2024automated}, directly relying on them for exploit generation is far from trivial due to two main challenges:

\begin{itemize}[leftmargin=*,noitemsep,topsep=5pt,parsep=0pt,partopsep=0pt]
    \item \textit{C1: How to precisely trigger a known vulnerability given heterogeneous and unstructured vulnerability information?}
Vulnerability information is scattered across diverse sources and expressed in mixed formats, including natural language explanations, HTTP request examples, payload fragments, and code snippets. These heterogeneous representations lack a unified structure, making it difficult to precisely trigger the vulnerability with correct technical details, including input data structure, parameter constraints, and vulnerability-critical characters. 

    \item \textit{C2: How to ensure reliable and precise exploit generation under unreliable LLM outputs?}
Large language models often produce outputs that contain incorrect assumptions at the level of concrete exploit details, making it difficult to adapt generated exploits to concrete deployment environments. 
Moreover, their outputs are highly inconsistent: given the same vulnerability information and attack objective, repeated generations frequently differ in low-level details, introducing significant uncertainty and further undermining the reliability of generated exploits.

\end{itemize}

Motivated by these, this paper aims to design a fully automated and reliable exploit generation approach that can effectively support known-vulnerability-based web application penetration testing. Specifically, we propose \tool, a multi-agent automated exploit generation framework for black-box web applications. Our design is motivated by the insight that vulnerability exploitation can be decomposed into structured trigger logic extraction and runtime adaptation. \tool consists of two main phases. In \textit{Phase A: Trigger Function Construction}, \tool extracts vulnerability trigger logic from heterogeneous and unstructured vulnerability information and encapsulates it into reusable trigger functions, providing a structured and consistent representation for downstream exploitation. In \textit{Phase B: Runtime Exploitation}, \tool instantiates trigger functions according to specific attack objectives and iteratively interacts with the target web application, refining exploits based on runtime feedback to adapt to the concrete execution environment.

We evaluate \tool on 104 real-world vulnerabilities deployed in concrete web application environments. We design 29 attack objectives, resulting in 660 concrete exploitation tasks. Across 55,440 exploit attempts, \tool achieves an average attack success rate of 82.41\%, substantially outperforming state-of-the-art baselines, whose best performance reaches only 32.88\%. Ablation studies further confirm the contribution of different agents in the framework.

In summary, the paper makes the following contributions:
\begin{itemize}[leftmargin=*,noitemsep,topsep=5pt,parsep=0pt,partopsep=0pt]
  \item We present \tool, a multi-agent framework for fully automated exploit generation targeting known vulnerabilities in black-box web applications.
  \item We propose a trigger function construction design that transforms unstructured vulnerability information into reusable, structured representations.
  \item We introduce a task decomposition and test-driven validation strategy that improves robustness and mitigates the impact of LLM hallucination.
  \item Our evaluation on 104 real-world vulnerabilities in web applications demonstrates the effectiveness of \tool.
\end{itemize}
\section{Problem Statement}

\begin{figure}[!t]
    \centering
    \includegraphics[width=\linewidth]{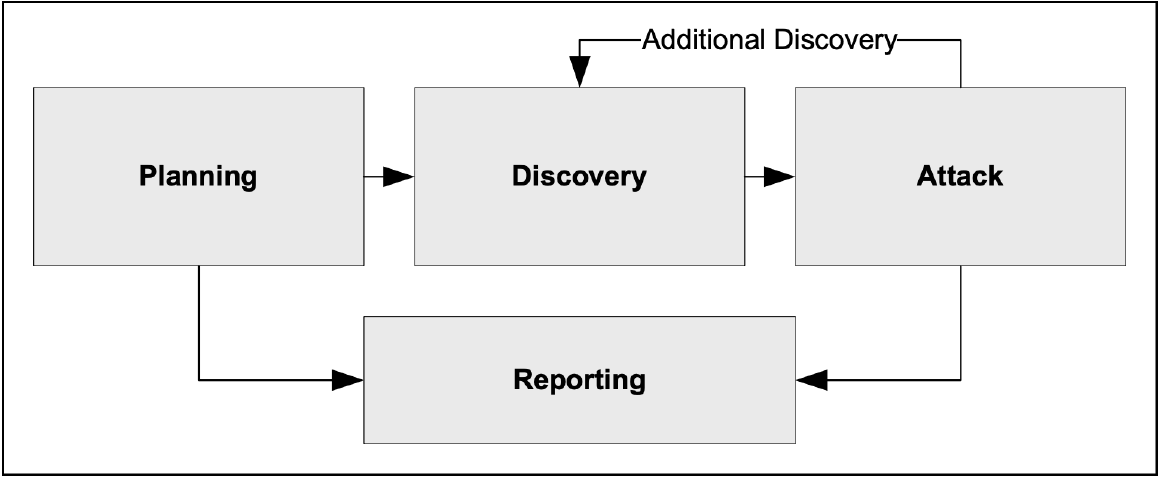}
    \caption{Four-stage penetration testing methodology defined by NIST (Figure~5-1), reproduced from NIST Special Publication~800-115~\cite{scarfone2008technical}.}
    \label{fig:nist}
\end{figure}

\subsection{Penetration Testing \& Exploit Generation}

According to the NIST technical guidance, \textit{penetration testing} follows a structured workflow consisting of four stages: \textit{Planning}, \textit{Discovery}, \textit{Attack}, and \textit{Reporting}~\cite{scarfone2008technical}, as illustrated in Figure~\ref{fig:nist}. 
The \textit{Planning} stage defines the testing scope, target assets, and attack objectives. 
The \textit{Discovery} stage then collects technical information about the target system, including exposed services, configurations, and publicly disclosed vulnerability information. 
Based on the data from previous stages, the \textit{Attack} stage performs concrete exploitation attempts against the running system to validate whether identified vulnerabilities can be successfully triggered and what practical security impact they produce. 
The \textit{Reporting} stage summarizes the observed results and security implications.

This paper focuses on the \textit{Attack} stage and studies \textit{exploit generation} as an independent technical task.
Concretely, the inputs include (i) vulnerability information, (ii) specified attack objectives, and (iii) access to a target web application. 
The output is an executable exploit that interacts with the target application through its exposed interfaces. 
A successful exploit must reliably trigger the known vulnerability and satisfy the specified attack objective under black-box conditions.

Our task differs substantially from proof-of-concept (PoC) generation or reproduction~\cite{simsek2025pocgen,marques2025automated,zhao2025systematic,mei2024arvo}. 
Such PoC-oriented tasks are typically studied under white-box settings with access to source code, whereas our exploit generation operates strictly under a black-box setting. 
PoC generation further focuses on new or previously unknown vulnerabilities, while both PoC reproduction and our task target \textit{known vulnerabilities}. 
However, PoC reproduction aims only to confirm the triggering condition, whereas our task targets full exploitation, requiring end-to-end executable exploits that achieve specified attack objectives in real-world web applications.

\subsection{Limitations of Vulnerability Information}
\label{sec:limitations-vuln-info}

Vulnerability information used for exploit generation is primarily obtained from public repositories such as \texttt{cve.org}~\cite{CVE} and the National Vulnerability Database (NVD)~\cite{nvd}. 
A typical record provides a unique identifier (e.g., CVE-2013-4547), a short textual description, and a set of external reference links. 
However, the descriptions are authored without standardized formats or explicit technical requirements for exploitation. 
They mainly summarize affected components, vulnerable versions, and high-level impact, while omitting many critical details required to construct a working exploit, including precise triggering conditions, request structures, parameter constraints, environmental assumptions, and dependency requirements~\cite{cve20195418,cve20201957,cve20259985}. 
As a result, vulnerability descriptions alone rarely provide sufficient information to directly derive executable exploits.

External references are intended to compensate for these missing details. 
However, only a small fraction of disclosed vulnerabilities are accompanied by publicly available exploit implementations~\cite{mu2018understanding,dang2025real,householder2020historical}; prior study reports that only 3,164 among 75,807 examined vulnerabilities have associated public exploits~\cite{householder2020historical}. 
Exploitation-related knowledge is instead scattered across heterogeneous and unstructured materials, spanning sources such as GitHub repositories, personal blogs, and security vendor reports. 
Available information often appears in fragmented forms, including \textit{natural-language explanations}, \textit{HTTP request examples}, \textit{payloads}, \textit{code snippets}, as well as hybrid combinations thereof.
Even when useful clues exist, they are rarely presented in a directly actionable or machine-consumable form and typically require substantial manual interpretation, normalization, and gap filling.

Consequently, translating the unstructured vulnerability information into reliable, executable exploits remains a fundamental challenge in real-world penetration testing. 
This difficulty stems from the need to simultaneously infer implicit triggering logic and operational constraints, and to concretely instantiate them into executable attack behaviors that remain robust under black-box interaction and environment-specific variability. 
Appendix~\ref{sec:appendix-vuln-examples} provides representative examples illustrating the diversity and unstructured nature of vulnerability information encountered in practice.

\subsection{Threat Model}

We consider an authorized attacker whose goal is to automatically generate working exploits for \textit{known vulnerabilities} against a target web application under black-box conditions. The attacker has no access to the application's source code, configuration files, or internal system state, and can interact with the target system only through its exposed HTTP(S) interfaces. The attacker is provided with publicly available vulnerability information, including CVE descriptions and the contents of their referenced links, as well as the target application URL and specified attack objectives that define the intended security impact. Based on these inputs, the attacker aims to generate executable exploits that can reliably trigger the known vulnerability and satisfy the specified objectives in the concrete deployment environment. We assume that vulnerability discovery, source-level analysis, and insider access are out of scope. Defensive evasion beyond the exploitation of the known vulnerability, post-exploitation activities, and lateral movement are also not considered in this work.




\subsection{A Real-World Exploit Generation Example}\label{sec:realexample}

\begin{figure}[!t]
    \centering
    \includegraphics[width=\linewidth]{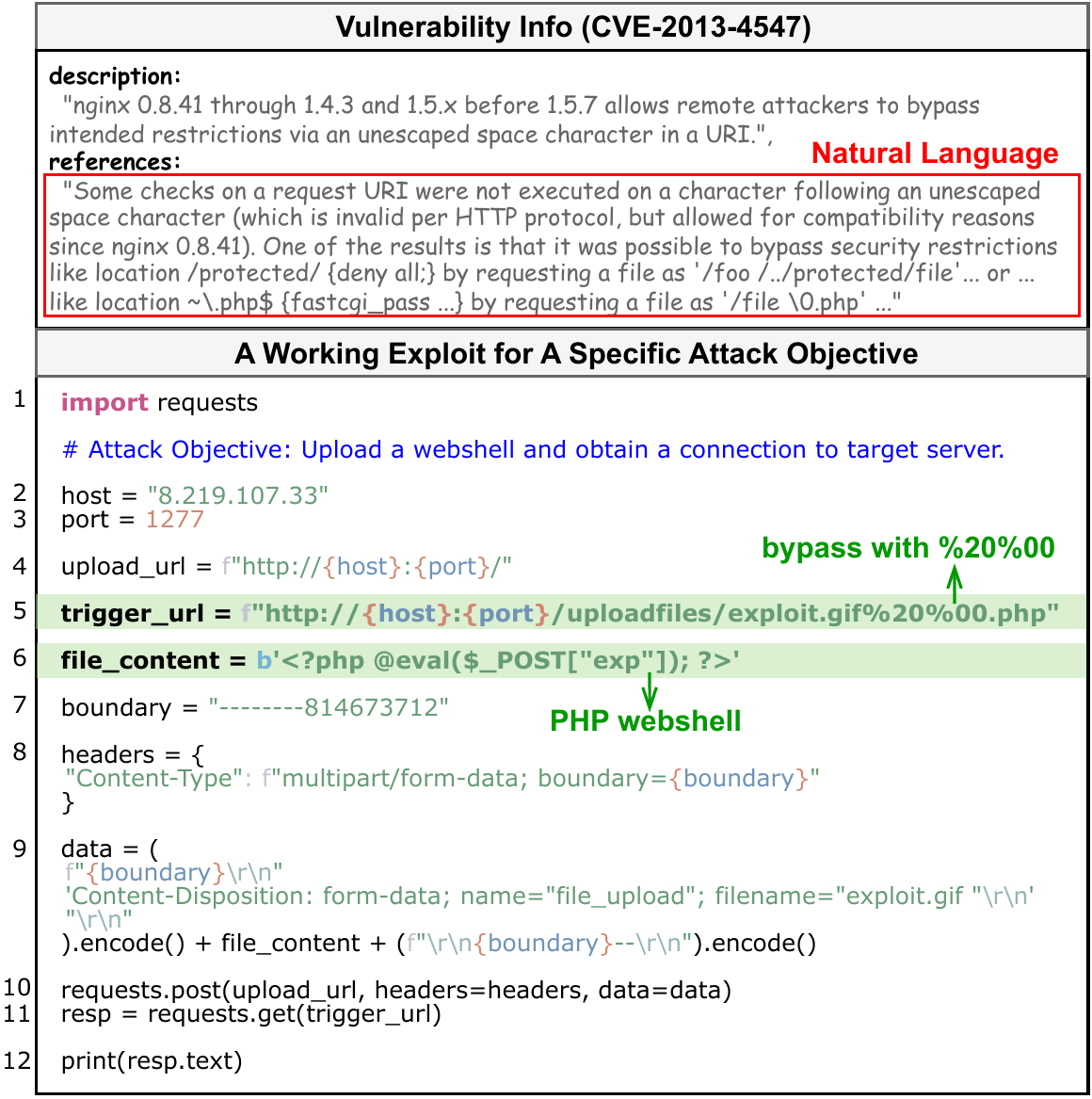}
    \caption{A real-world exploit generation example based on CVE-2013-4547.}
    \label{fig:real-example}
\end{figure}

We illustrate the practical challenges of exploit generation through a real-world example based on CVE-2013-4547~\cite{cve20134547}, a canonical Nginx URI parsing bypass vulnerability that enables access control evasion via crafted request paths. This vulnerability remains widely present in today’s web applications, as large-scale Internet measurements from platforms such as Shodan and Censys continue to observe thousands of publicly reachable services affected~\cite{shodandata,shodanexample,censys}. Moreover, it can be exploited to achieve diverse attack objectives, including unauthorized access, remote command execution, and webshell upload, making it a representative and high-impact case for studying real-world exploit generation.

Figure~\ref{fig:real-example} presents the vulnerability information together with the generated exploit details for this example. The references linked from the CVE entry~\cite{cve20134547reference} mainly consist of natural language descriptions outlining high-level bypass strategies, such as injecting special characters into request paths, including traversal-like patterns and null-byte-related encodings, rather than a concrete exploit script. In this example, the attack objective is to \textit{upload a webshell to the target application}, which requires constructing a precise HTTP request that simultaneously satisfies multiple constraints. Specifically, the request must embed a correctly encoded bypass payload at the appropriate position in the request path (line 5) and include a well-formed malicious file whose content constitutes a valid executable webshell (line 6). A working exploit depends on the joint correctness of the bypass payload, the HTTP request structure, and the webshell content. Even a single incorrect character in the crafted path or payload can invalidate the request and cause the vulnerability trigger to fail. This example demonstrates the substantial gap between high-level vulnerability information and a fully working exploit, highlighting that exploit generation in real-world penetration testing is a complex and error-prone task.

\section{Related Work}

\subsection{White-Box Vulnerability Detection}
Existing vulnerability detection techniques are predominantly designed for white-box settings, where full access to the target application's source code is assumed~\cite{wu2018fuze,manes2019art,baldoni2018survey,ryan2023sylvia}. These techniques aim to detect and validate previously unknown vulnerabilities through program analysis methods such as fuzzing and symbolic execution. In contrast, known-vulnerability-based penetration testing focuses on exploiting already disclosed vulnerabilities and is typically conducted under black-box conditions, where source code is unavailable. As a result, white-box vulnerability detection techniques are not applicable to the problem setting considered in this work and are therefore excluded from our scope.

\subsection{Automated Penetration Testing with LLMs}\label{sec:relatedllmworks}

Recent approaches employ large language models to automate penetration testing workflows~\cite{pentestgpt,shen2025pentestagent,kong2025vulnbotautonomouspenetrationtesting,huang2023penheal,xu2024autoattacker,happe2023getting}. These systems typically include an exploitation component that generates attack scripts for known vulnerabilities. In most existing designs, exploit generation is driven by directly prompting the model to translate vulnerability information into executable exploits, without introducing explicit intermediate representations, constraints or systematic validation.

Under this general paradigm, representative systems mainly differ in how they organize the interaction between the LLM and exploit execution. PentestGPT~\cite{pentestgpt} adopts a chatbot-style interaction model, where the LLM generates exploit scripts based on conversational context, while execution and result interpretation are performed by the user and fed back into the dialogue. PentestAgent~\cite{shen2025pentestagent} introduces an agent-based workflow where exploit generation is guided by two predefined questions prompting the LLM to identify execution parameters and other required information. The generated exploits are executed by a dedicated component. VulnBot~\cite{kong2025vulnbotautonomouspenetrationtesting} further extends this design by employing multiple agents, where a generator produces exploits from vulnerability scan results and an executor iteratively runs them toward a specified attack objective. Despite these additional designs, exploit generation in all three systems ultimately relies on the LLM to directly produce complete exploits in a single step, without explicit verification or constraint over intermediate variables such as parameters, payload structures, or vulnerability trigger conditions. The generated exploits often omit critical parameters, alter vulnerability-sensitive characters, or reference non-existent environment-specific elements~\cite{yang2025pentesteval}, leading to unstable behavior across executions and attack objectives.

\section{Design of \tool}

\subsection{Challenges \& Insights}\label{sec:challenges}
Existing works (in Section~\ref{sec:relatedllmworks}) demonstrate that LLMs hold promise for exploit generation. However, directly applying LLMs to this task remains of limited effectiveness in practice~\cite{yang2025pentesteval}. Two main challenges must be carefully addressed.

\noindent\textbf{Challenge I: How to precisely trigger a known vulnerability given heterogeneous and unstructured vulnerability information?}
The complexity of vulnerability descriptions makes precise triggering difficult, as critical constraints and dependencies are often implied rather than explicitly stated. In the example presented in Section~\ref{sec:realexample}, the references consist solely of natural-language explanations, requiring the attacker to infer subtle triggering details before constructing a concrete payload and HTTP request. This illustrates that identifying vulnerability-critical inputs and structural constraints is non-trivial for practical exploit generation. Existing LLMs frequently miss essential parameters or incorrectly modify vulnerability-sensitive symbols, such as encoded strings or special delimiters, which breaks the triggering behavior.

\underline{\textit{Solution:}}
To address this challenge, we decouple vulnerability triggering reasoning from exploit generation as an independent abstraction step. Under our threat model, attackers interact with the target application solely through its exposed HTTP(S) interface, implying that successful exploitation ultimately reduces to constructing vulnerability-specific HTTP requests. Since HTTP interactions are inherently structured, we propose a \textbf{\textit{trigger function abstraction}} that captures the inferred triggering semantics as an instantiable functional representation over HTTP requests. The abstraction consolidates invariant elements (e.g., fixed endpoints, required parameters, and constant fields) while exposing only the minimal set of variable components as configurable inputs. This design converts implicit triggering knowledge into a unified executable form, enabling reliable downstream invocation, systematic validation, and reuse across multiple exploit realizations.

\noindent\textbf{Challenge II: How to ensure reliable and precise exploit generation under unreliable LLM outputs?}
Exploit generation requires producing concrete attack codes that can be reliably executed against a target web application, where even minor deviations may invalidate the attack. 
However, LLM-generated exploits often contain invalid elements, such as non-existent paths, files, or API endpoints, leading to execution failures in black-box testing. 
Beyond explicit errors, output instability poses a deeper challenge: even with identical vulnerability information and attack objectives, LLMs frequently generate exploits that differ in low-level details such as parameter selection and argument ordering. 
This variability makes exploit generation difficult to control and reason about. 
When exploit generation is extended to support multiple attack objectives for the same vulnerability, such instability persists across parallel exploit realizations, significantly undermining the reliability and practicality of LLM-based approaches.

\underline{\textit{Solution:}}
To mitigate the impact of unreliable LLM outputs, we \textit{decompose exploit generation} into sub-tasks and introduce \textit{\textbf{test-driven validation}} to constrain each intermediate artifact. 
Rather than relying on a monolithic end-to-end generation process, each sub-task produces an intermediate result with explicit structural or semantic expectations that can be verified before being consumed by other components. 
By constraining and validating intermediate artifacts, this design limits the introduction and propagation of incorrect or inconsistent elements during exploit construction. 
In addition, the trigger function introduced in \textit{Challenge~I} further stabilizes exploit generation by reducing free-form code synthesis to structured function instantiation, which inherently narrows the output space and reduces opportunities for incorrect or inconsistent elements. 
Together, these designs improve the reliability and controllability of exploit generation, even when supporting multiple attack objectives for the same vulnerability.

\begin{figure*}[!t]
    \centering
    \includegraphics[width=0.9\linewidth]{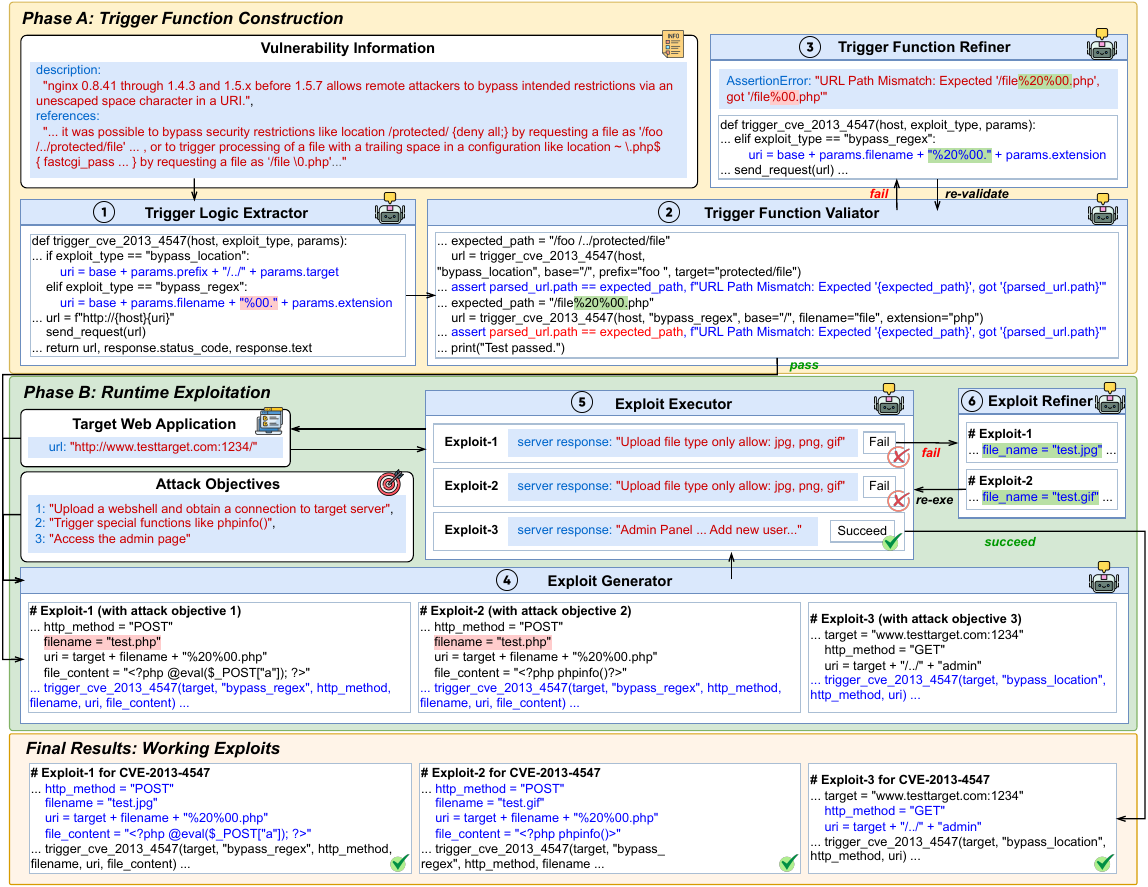}
    \caption{Overall architecture of \tool with a running example (CVE-2013-4547).}
    \label{fig:overview}
\end{figure*}

\subsection{Overview}
Building on the above insights, we present \tool, a multi-agent framework for automatic exploit generation targeting known vulnerabilities in black-box web applications. Figure~\ref{fig:overview} illustrates the overall architecture of \tool using a running example based on CVE-2013-4547. The framework organizes exploit generation into two complementary phases that enable precise and reliable exploitation:

\begin{itemize}[leftmargin=*,noitemsep,topsep=5pt,parsep=0pt,partopsep=0pt]
  \item \textit{Phase A: Trigger Function Construction (\S\ref{sec:phase1})} extracts precise vulnerability trigger logic from heterogeneous and unstructured vulnerability information and produces reusable trigger functions. This phase is realized through the collaboration of three agents: \textit{\agent{1} Trigger Logic Extractor} derives trigger logic from vulnerability information and constructs initial trigger functions; \textit{\agent{2} Trigger Function Validator} evaluates the correctness of the generated trigger functions; and \textit{\agent{3} Trigger Function Refiner} revises trigger functions that fail validation. The refined trigger functions are fed back to the validator, enabling repeated interaction between \agent{2} and \agent{3} until a valid trigger function is obtained or a predefined revision limit is reached.

  \item \textit{Phase B: Runtime Exploitation (\S\ref{sec:phase2})} builds on the trigger function produced in \textit{Phase~A} to instantiate and execute concrete exploit instances for different attack objectives against the target web application. This phase is realized through the collaboration of three agents: \textit{\agent{4} Exploit Generator}, which instantiates the trigger function into an exploit instance for each specified attack objective; \textit{\agent{5} Exploit Executor}, which executes each exploit instance, collects runtime feedback (e.g., server responses) from the target application, and determines whether the attack is successful; and \textit{\agent{6} Exploit Refiner}, which revises exploit instances that fail execution based on the observed execution results. Each revised exploit is returned to the executor for re-execution, enabling repeated interaction between \agent{5} and \agent{6} until the corresponding attack objective is achieved or a predefined revision limit is reached.
\end{itemize}

\noindent For reproducibility and transparency, Appendix~\ref{sec:appendix-prompts} provides the complete prompts used by each agent in the framework.

\subsection{Phase A: Trigger Function Construction}\label{sec:phase1}
In this phase, \tool focuses on extracting precise vulnerability trigger logic from heterogeneous vulnerability information, including CVE descriptions and reference materials, and formalizing it into a reliable \textit{trigger function}. By decoupling trigger logic construction from exploit generation, \textit{Phase~A} transforms unstructured vulnerability knowledge into a reusable and executable abstraction, which serves as the foundation for subsequent runtime exploitation.

\subsubsection{\agent{1} Trigger Logic Extractor}\label{sec:trigger-extractor} 

\begin{figure}[!t]
    \centering
    \includegraphics[width=0.85\linewidth]{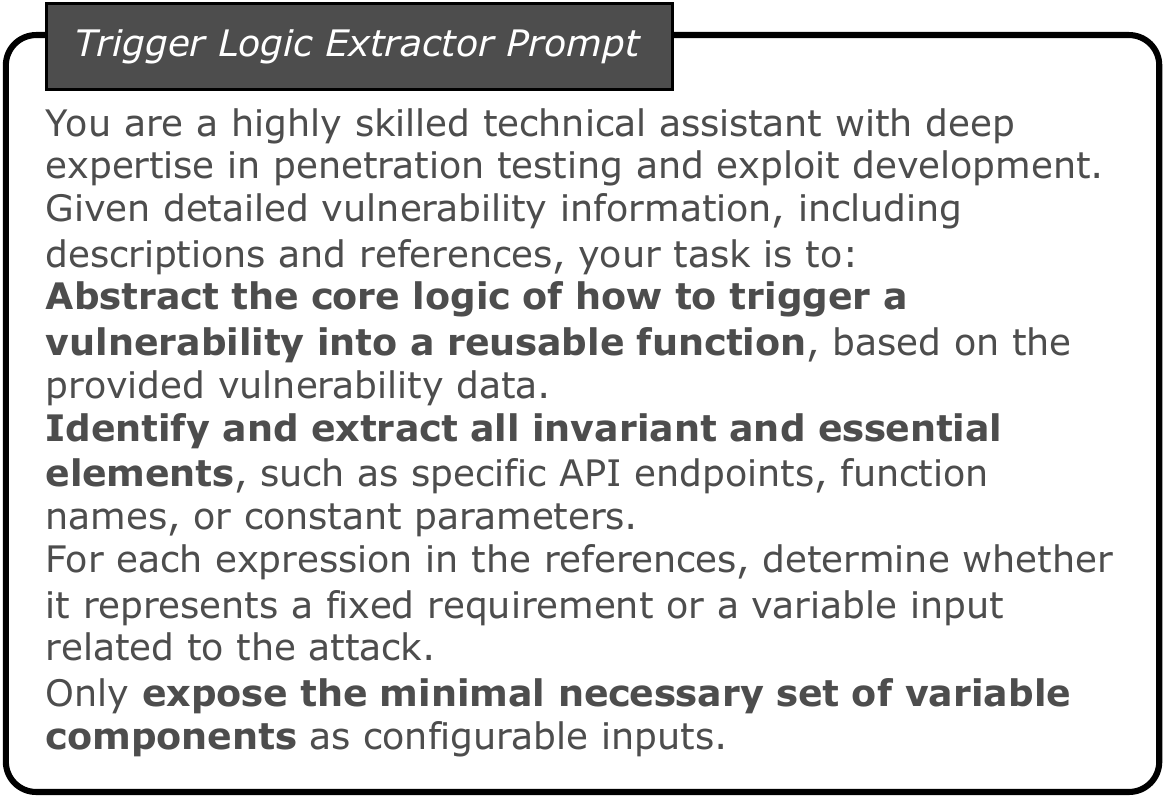}
    \caption{Prompt in the \textit{Trigger Logic Extractor} agent.}
    \label{fig:extractor-prompt}
\end{figure}

This agent extracts vulnerability trigger logic from unstructured vulnerability information and formalizes it as an initial trigger function. Given CVE descriptions and reference materials, it identifies the core logic required to trigger the vulnerability and encodes it into a reusable functional representation. Logic that must remain unchanged to preserve the vulnerability semantics is embedded directly into the trigger function, while only a small and well-defined set of elements that vary across exploit instances is exposed as configurable inputs. By doing so, the \textit{Trigger Logic Extractor} explicitly separates vulnerability semantics from exploit instantiation choices, ensuring that all subsequent exploit generation follows a consistent trigger logic rather than reinterpreting the vulnerability. Within the framework, the trigger function produced by this agent serves as the only interface through which later agents access vulnerability semantics. The extraction process is guided by a structured prompt, shown in Figure~\ref{fig:extractor-prompt}.

Using the running example (in Figure~\ref{fig:overview}), we illustrate how the \textit{Trigger Logic Extractor} operates when vulnerability information is provided entirely in natural language. In CVE-2013-4547, the description states that unescaped space characters in Nginx URI parsing allow attackers to bypass access restrictions, and the references describe two common bypass patterns: path-based location bypass and regex-based suffix bypass. Based on this information, the extractor captures the unescaped space handling as shared trigger logic and represents the two bypass patterns as fixed branches within the trigger function. Inputs such as path prefixes, target resources, filenames, or extensions are treated as configurable, while the placement and role of special characters remain fixed. As a result, different exploit instances targeting different attack objectives can be derived from the same trigger function without reinterpreting the underlying vulnerability logic.

\subsubsection{\agent{2} Trigger Function Validator}\label{sec:trigger-validator}

This agent determines whether the trigger function produced by \agent{1} correctly reflects the vulnerability trigger logic described in the vulnerability information. Instead of relying on free-form reasoning, this agent performs validation through a test-driven and rule-based process. Specifically, it automatically generates a structured test case from the vulnerability references using an LLM, and then executes the test case against the trigger function using deterministic checks implemented as assertion statements. Each test case specifies the expected HTTP semantics implied by the references, including the URL path (and query parameters when applicable), the HTTP request body, and the HTTP method or headers that are critical for successful exploitation. The validation process outputs \texttt{``Test passed.''} if all checks succeed; otherwise, it raises an \texttt{AssertionError} with a clear and localized error message indicating the mismatched component.

Using the running example (in Figure~\ref{fig:overview}), we illustrate how the \textit{Trigger Function Validator} detects subtle but vulnerability-critical errors that are difficult to identify during trigger logic extraction alone. For CVE-2013-4547, the references describe bypass requests whose URL paths must contain special characters at precise positions. Accordingly, the generated test case instantiates the trigger function under different bypass modes and asserts an exact match between the expected URL path derived from the references and the actual path produced by the trigger function. As shown in the example, validation fails when the trigger function incorrectly constructs the path \texttt{/file\%00.php} for the regex-based bypass, whereas the reference-consistent path should be \texttt{/file\%20\%00.php}. The resulting assertion pinpoints the missing encoded space and returns the failing trigger function unchanged, providing concrete feedback for subsequent revision. By enforcing strict consistency with the vulnerability references, the \textit{Trigger Function Validator} prevents subtle symbol-level errors from propagating into downstream exploit generation.

\subsubsection{\agent{3} Trigger Function Refiner}\label{sec:trigger-refiner}

When \agent{2} reports a validation failure, this agent revises the trigger function based on the corresponding \texttt{AssertionError}. It takes as input the failing trigger function together with this \texttt{AssertionError}, which explicitly identifies the mismatch between the expected behavior derived from the vulnerability references and the actual behavior produced by the trigger function. Rather than regenerating trigger logic from scratch, the refiner performs targeted revisions that address the reported error while preserving the original intent, structure, and vulnerability semantics encoded in the trigger function. In this way, refinement is driven by concrete validation signals and remains focused on correcting specific inconsistencies, avoiding unnecessary or disruptive changes.

Using the running example (in Figure~\ref{fig:overview}), we illustrate how the \textit{Trigger Function Refiner} corrects vulnerability-critical errors exposed during validation. For CVE-2013-4547, the validator reports an error (i.e., \texttt{URL Path Mismatch: Expected `/file\%20\%00.php', got `/file\%00.php'}), indicating that the generated trigger function omits an encoded space character required by the reference-consistent bypass request. Guided by this \texttt{AssertionError}, the refiner updates the corresponding branch of the trigger function by restoring the missing encoded space in the constructed URI (i.e., revising \texttt{``\%00.''} to \texttt{``\%20\%00.''}). The refined trigger function is then returned to \agent{2} for re-validation. This refinement--validation loop continues until the trigger function passes all checks or a predefined revision limit is reached (three iterations in our implementation). By coupling precise assertion feedback with bounded refinement, the \textit{Trigger Function Refiner} incrementally improves trigger function correctness and ensures that only validated trigger logic is propagated to downstream exploit generation.

\subsection{Phase B: Runtime Exploitation}\label{sec:phase2}

In this phase, \tool focuses on instantiating and executing concrete exploit instances for different attack objectives against the target web application, based on the validated trigger function produced in \textit{Phase~A}. By grounding exploit generation in verified trigger logic, \textit{Phase~B} avoids reinterpreting vulnerability information and instead focuses on practical runtime exploit execution under the black-box conditions.

\subsubsection{\agent{4} Exploit Generator}\label{sec:exploit-generator}

This agent instantiates the validated trigger function into concrete, executable exploits for different attack objectives. It takes as input the trigger function produced in \textit{Phase~A}, the target web application (URL), and a set of attack objectives, and generates one standalone exploit instance for each objective. Concretely, this agent is responsible for binding the abstract trigger logic to a specific target environment by selecting concrete parameter values, constructing complete HTTP request flows, and embedding the trigger function into an executable script. By design, the trigger function remains unchanged across all generated exploits and serves as the sole carrier of vulnerability semantics, while differences among exploit instances are introduced only through objective-specific parameters and payload construction. This separation ensures exploit diversity through controlled instantiation rather than reinterpreting or modifying the underlying vulnerability logic.

Using the running example (in Figure~\ref{fig:overview}), we illustrate how the \textit{Exploit Generator} derives multiple exploit instances from a single trigger function for CVE-2013-4547. Given the same validated trigger function and target URL, the agent generates three exploits corresponding to distinct attack objectives: uploading a webshell, triggering diagnostic functionality (e.g., \texttt{phpinfo()}), and accessing an administrative endpoint. For the first two objectives, the trigger function is instantiated in the regex-based bypass mode with different payload contents, resulting in exploits that share identical bypass logic but differ in injected code. For the third objective, the trigger function is instantiated in the path-based bypass mode to construct a traversal request targeting the administrative path. In all cases, the trigger function itself is reused without modification, and objective-specific behavior is realized solely through controlled parameter instantiation.

\begin{figure}[!t]
    \centering
    \includegraphics[width=0.85\linewidth]{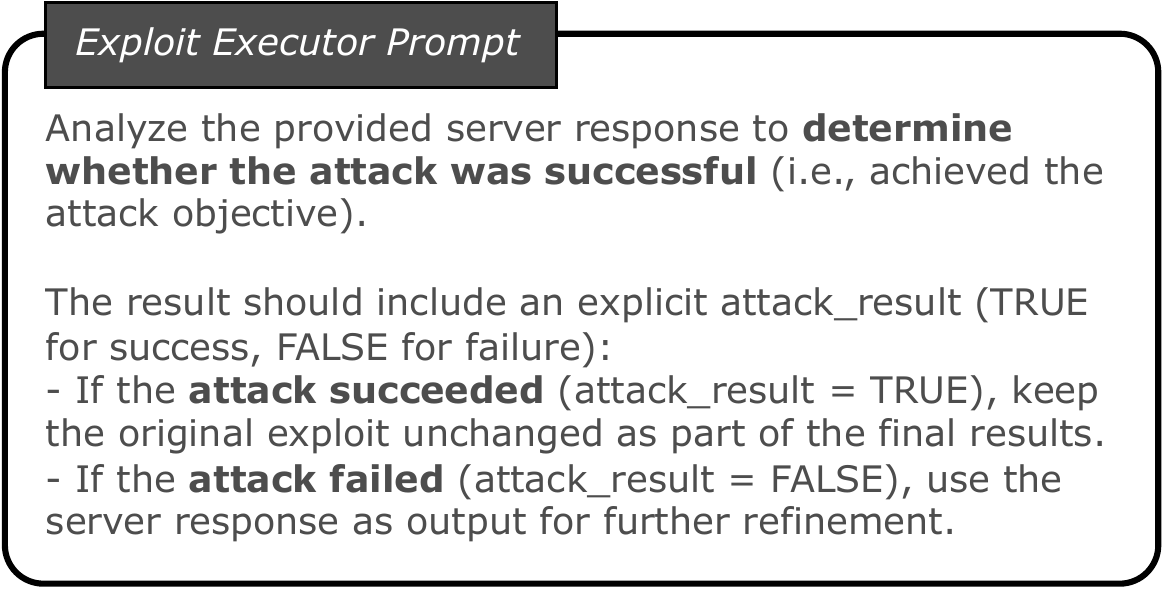}
    \caption{Prompt in the \textit{Exploit Executor} agent.}
    \label{fig:executor-prompt}
\end{figure}

\subsubsection{\agent{5} Exploit Executor}\label{sec:exploit-executor}

This agent executes the exploits generated by \agent{4} against the target web application and determines whether each attack objective is achieved. It takes as input a set of exploit scripts together with their corresponding attack objectives, runs each exploit in the target environment, and collects execution feedback such as server responses. For each exploit, the agent produces an explicit success or failure decision based on whether the observed runtime behavior satisfies the specified attack objective. To support this decision making, the \textit{Exploit Executor} employs an LLM-guided assessment that analyzes server responses with respect to the attack objective and outputs a binary \texttt{attack\_result}, as constrained by the prompt shown in Figure~\ref{fig:executor-prompt}. Successful exploits are directly recorded as final results, while failed exploits are returned together with their execution feedback for subsequent refinement.

Using the running example (in Figure~\ref{fig:overview}), three exploits are executed targeting webshell upload, diagnostic code execution and admin panel access respectively for CVE-2013-4547. The first two exploits are rejected by the server due to file-type restrictions, and are therefore marked as failed, with their server responses preserved as feedback. In contrast, the third exploit successfully retrieves the admin interface and is marked as successful. This separation ensures that only exploits achieving concrete security impact are retained, while informative failure feedback is propagated to the refinement stage.

\subsubsection{\agent{6} Exploit Refiner}\label{sec:exploit-refiner}

This agent revises failed exploits to better satisfy the attack objectives under the constraints of target environment. It operates on a failed exploit together with the corresponding server response produced by \agent{5} and performs targeted modifications that address the cause of the failure while preserving the original attack objective and trigger function. Refinement focuses on identifying which components are rejected at runtime and selectively adjusting only the minimal elements (e.g., parameter values, payload formats, or request construction details) necessary to resolve the failure. Refined exploits are returned to \agent{5} for re-execution, forming a bounded refinement loop that continues until the attack succeeds or a predefined revision limit is reached (three iterations in our implementation).

Using the running example shown in Figure~\ref{fig:overview}, we illustrate how the \textit{Exploit Refiner} adapts exploits to environment-specific constraints for CVE-2013-4547. For the first two attack objectives, \agent{5} reports failure with server responses indicating that only image file types (e.g., \texttt{jpg}, \texttt{png}, \texttt{gif}) are allowed for upload. Guided by this feedback and the original objectives, the refiner revises the corresponding exploit scripts by modifying the uploaded file names to compliant formats (e.g., from \texttt{``test.php''} to \texttt{``test.jpg''} or \texttt{``test.gif''}), while leaving the trigger function and overall exploit structure unchanged. The refined exploits are then returned to \agent{5} for re-execution. Through this feedback-driven refinement process, exploits are incrementally adapted to practical deployment constraints without reinterpreting the trigger function or altering the intended attack objectives.

\section{Experiments}\label{sec:experiments}

We evaluated \tool on an established benchmark to demonstrate its effectiveness on exploit generation. Specifically, we address the following three research questions:

\begin{itemize}[leftmargin=*,noitemsep,topsep=5pt,parsep=0pt,partopsep=0pt]
\item \textbf{RQ1 (\tool Performance)}: How does \tool perform on exploit generation for real-world vulnerabilities?

\item \textbf{RQ2 (Comparative Effectiveness)}: How does \tool perform compared to existing methods?  

\item \textbf{RQ3 (Ablation Study)}: What is the individual contribution of each core innovation to the overall performance?  
\end{itemize}

\begin{table*}[t]
    \caption{Details of the seven objective categories, with predefined attack objective sets and related CVE identifiers.}
    \label{tab:cve-s}
    \centering
    \resizebox{\textwidth}{!}{
    \begin{tabular}{c|l|c}
    \toprule
    \textbf{Categories} & \textbf{Attack Objective Set} & \textbf{Related CVEs}\\
    \midrule
    \makecell{\textbf{Sensitive Information (\textit{SI})}} & \makecell[l]{
\textit{- Exfiltrate the content of WEB-INF/web.xml} 
\\ \textit{- Read the contents of /etc/passwd file from the target server ...} 
} & 
\makecell[l]{CVE-2018-3760, CVE-2024-4956, ... (\textbf{24} CVEs)}\\
    \midrule

    \makecell{\textbf{Remote Command Execution (\textit{RCE})}} & \makecell[l]{\textit{- Get information about the UID and GID settings on the target server}
\\ \textit{- Identify the current user on target server during the remote command execution ...} 
} & 
\makecell[l]{CVE-2019-17558, CVE-2023-26360 ... (\textbf{58} CVEs)}  \\
    \midrule
    
    \makecell{\textbf{Arbitrary File Write (\textit{AFW})}} & \makecell[l]{\textit{- Write '1234567890' into the file /tmp/test0001 remotely}
\\ \textit{- Create a new file success.txt under the current path remotely ...}
} & 
\makecell[l]{CVE-2023-38646, CVE-2020-13942 ... (\textbf{27} CVEs)}  \\
    \midrule
    
    \makecell{\textbf{Privilege Escalation (\textit{PE})}} & \makecell[l]{
\textit{- Add a system account(testA) and show all accounts on the server from /etc/passwd ...}
} & 
\makecell[l]{CVE-2023-32315, CVE-2024-43441 ... (\textbf{9} CVEs)}  \\
    \midrule
    
    \makecell{\textbf{Webshell (\textit{WS})}} & \makecell[l]{\textit{- Upload a webshell and obtain a connection to target server}
\\ \textit{- Obtain the reverse shell connection ...} 
} & 
\makecell[l]{CVE-2017-10271, CVE-2017-9841 ... (\textbf{5} CVEs)}  \\
    \midrule
    
    \makecell{\textbf{SQL Injection (\textit{SQLi})}} & \makecell[l]{
\textit{- Obtain table or schema data via SQL injection ...}} & 
\makecell[l]{CVE-2020-9402, CVE-2023-25157 ... (\textbf{8} CVEs)}  \\
    \midrule
    
    \makecell{\textbf{Miscellaneous (\textit{MISC})}} & \makecell[l]{
\textit{- Redirect user to `www.google.com` ...} 
} & 
\makecell[l]{CVE-2021-40822, CVE-2021-40438 ... (\textbf{4} CVEs)}  \\
\bottomrule
\end{tabular}
}
\end{table*}

\subsection{Experimental Setup}\label{sec:experiment-setting}

\subsubsection{Benchmark Setup}
We derive a dataset of 104 web-application vulnerabilities by filtering the benchmark used in \textit{PentestAgent}~\cite{pentestgpt}, excluding non-web entries such as Linux kernel issues. For each vulnerability, we use the corresponding vulnerable application environment provided by Vulhub~\cite{vulhub}, where the vulnerability can be triggered in a real application context. Each environment is deployed as an isolated Docker instance to ensure independence and reproducibility. The dataset reflects vulnerabilities commonly encountered in real-world web applications and covers 18 of the top 25 most dangerous software weaknesses in the Common Weakness Enumeration (CWE) list~\cite{CWE}.

\subsubsection{Vulnerability Information Dataset}
As discussed in Section~\ref{sec:limitations-vuln-info}, vulnerability information for exploit generation is heterogeneous and unstructured. 
The information comprises a description and a set of reference materials. Since CVE descriptions are consistently provided as natural language summaries, they are not further categorized. Instead, we categorize the 104 vulnerabilities into seven information types based on the content form of its references: \textbf{Natural Language (\textit{NL})}, \textbf{HTTP Request Examples (\textit{HTTP})}, \textbf{Payloads (\textit{Payload})}, \textbf{Code Snippets (\textit{Code})}, and three hybrid forms, \textbf{\textit{NL+HTTP}}, \textbf{\textit{NL+Payload}}, and \textbf{\textit{NL+Code}}. 
The dataset is distributed as follows: \textit{NL} (5), \textit{HTTP} (37), \textit{Payload} (17), \textit{Code} (3), \textit{NL+HTTP} (32), \textit{NL+Payload} (7), and \textit{NL+Code} (3), as details provided in Appendix~\ref{sec:appendix-taxonomy}.
These categories are used in RQ1 (\S\ref{sec:rq1}) to analyze how information types affect phase-wise and end-to-end exploit generation performance.

\subsubsection{Attack Objective Dataset}
For each vulnerability and its corresponding target web application deployed in the Docker environment, our penetration experts analyze feasible attack vectors, validate them through execution, and derive concrete attack objectives achievable in the environment; a single vulnerability may yield multiple distinct objectives. We then organize the collected objectives into seven categories guided by established security standards, including CWE~\cite{CWE} and OWASP Top 10~\cite{OWASP}. An exploit is considered successful if it achieves a specific annotated attack objective. Table~\ref{tab:cve-s} summarizes the objective categories, contents, and the number of vulnerabilities associated with each category, while the full mapping is provided in Appendix~\ref{sec:appendix-taxonomy}. Seven categories used in RQ2 (\S\ref{sec:rq2}) are defined as follows:

\begin{itemize}[leftmargin=*,noitemsep,topsep=5pt,parsep=0pt,partopsep=0pt]
\item \textbf{Sensitive Information (\textit{SI})}: Extracting sensitive data from the target system, such as credential files, configuration files, or environment variables.

\item \textbf{Remote Command Execution (\textit{RCE})}: Executing arbitrary commands on the target system through mechanisms such as command injection, deserialization, or evil router usage.

\item \textbf{Arbitrary File Write (\textit{AFW})}: Writing files to arbitrary locations on the server, potentially enabling data corruption, persistence, or further exploitation.

\item \textbf{Privilege Escalation (\textit{PE})}: Elevating privileges from the web application layer to higher-privileged system access.

\item \textbf{Webshell (\textit{WS})}: Deploying a persistent or interactive webshell on the server, such as uploading a backdoor file or establishing a reverse shell.

\item \textbf{SQL Injection (\textit{SQLi})}: Exploiting SQL injection vulnerabilities to achieve unauthorized data access or command execution via database-linked interfaces.

\item \textbf{Miscellaneous (\textit{MISC})}: Less frequent but impactful objectives involving auxiliary behaviors, such as \textit{SSRF}, \textit{XSS}, \textit{CSRF}, and \textit{URL redirection}.
\end{itemize}

\subsubsection{Large Language Models.} We evaluate \tool with four representative models: \textit{Qwen-Plus} (\texttt{qwen-plus-2025-04-28}), \textit{GPT-4o} (\texttt{gpt-4o-2024-08-06}), \textit{DeepSeek-V3} (\texttt{DeepSeek-V3-0324}) and \textit{Claude-3.7} (\texttt{claude-3-7-sonnet-20250219})~\cite{qwen3,gpt,deepseek,claude}.
All models are accessed through official APIs without prompt tuning or instruction fine-tuning. Default decoding parameters are used for all completions, including temperature, top-p, top-k (if applicable), and max token limits as set by provider. Input prompts are formatted uniformly across models.

\begin{figure}[!t]
    \centering
    \includegraphics[width=0.85\linewidth]{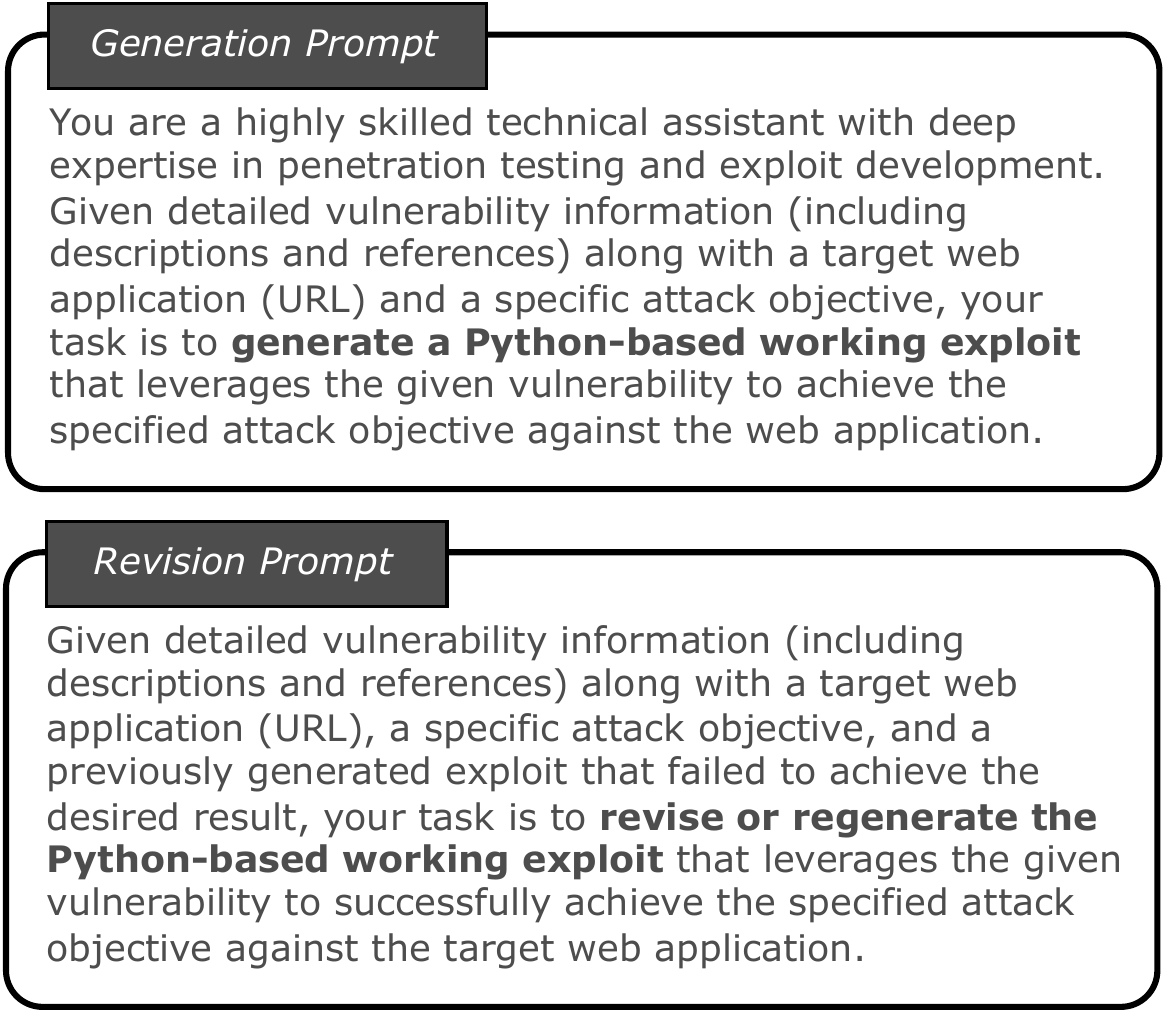}
    \caption{Prompts employed by \textit{DireLLM}.}
    \label{fig:direllm}
\end{figure}

\subsubsection{Baseline Methods.}
\label{sec:baselines}

We compare \tool against four representative automated penetration testing methods. To ensure a fair comparison, all methods are evaluated under consistent settings: receiving identical inputs (including the vulnerability information and predefined attack objectives) and operating against same target environments with same LLM backbones. During evaluation, each method is constrained by a unified retry budget of three attempts for each attack objective. 
For methods that do not impose an explicit budget configuration, such as \textit{PentestGPT} and \textit{VulnBot}, which rely on their own decision-making modules to determine termination, we modified their implementations and created variants adopting same retry limit as \tool for a fair comparison.
Baseline details are as follows:

\begin{itemize}[leftmargin=*]

\item \textit{Direct LLM Generation (\textit{DireLLM})}: 
This baseline directly prompts the LLM to generate an executable exploit from inputs (Figure~\ref{fig:direllm}). The generated exploit is executed and iteratively revised, following the same retry budget as \tool. This baseline represents the common practice of treating exploit generation as a direct text-to-code generation task without explicit intermediate representations or constraints.

\item \textit{PentestAgent}~\cite{shen2025pentestagent}: 
We adopt the official implementation and default configuration of \textit{PentestAgent}. Its exploit execution loop is constrained to the same retry budget as \tool to ensure comparable execution opportunities.

\item \textit{PentestGPT}~\cite{pentestgpt}: 
Since \textit{PentestGPT} is designed as a semi-automated, user-in-the-loop system, we implement an automated variant by replacing the human user with an execution agent, denoted as \textit{PentestGPT-Auto}. The agent executes generated exploits, captures server responses, and feeds them back into the dialogue to guide subsequent interactions. In addition to its default termination behavior, we configure a budget-constrained variant, \textit{PentestGPT-Auto-3}, which adopts the same retry budget as \tool.

\item \textit{VulnBot}~\cite{kong2025vulnbotautonomouspenetrationtesting}. 
We use the original multi-agent implementation of \textit{VulnBot} and preserve its internal control logic for iterative exploit generation and execution. Similar to \textit{PentestGPT-Auto}, we additionally configure a budget-constrained variant, \textit{VulnBot-3}, which operates under the same retry budget as \tool.

\end{itemize}

\subsubsection{Evaluation Metrics.}
Following the setting of PentestAgent~\cite{shen2025pentestagent}, we adopt the Attack Success Rate (ASR) as the evaluation metric. Specifically, given a set of CVEs $C$, its ASR is defined as:

\begin{equation}
\text{ASR}(C) = 
\frac{\sum_{c \in C} \left|\{\, o \mid o \in OBJ(c) \wedge success(c, o)\,\}\right|}
{\sum_{c \in C} \left|\{\, o \mid o \in OBJ(c)\,\}\right|}
\end{equation}

where  $OBJ(c)$ is the set of attack objectives for a given vulnerability $c$, $success(c, o)$ indicates that objective $o\in OBJ(c)$ for the CVE $c$ is successfully achieved by a given tool.
We can also compute ASR for a specific category. For example,
ASR($C_{SI}$) denotes the ASR for the SI category, where $C_{SI}$ is the set of 24 CVEs in that category.

\subsubsection{Implementation and Configuration.}
We implement \tool using a Python-based framework that interfaces with multiple LLM backends and evaluates generated exploits against the constructed benchmark. All methods, including \tool and baselines, are executed under a unified retry budget, where each iterative step is limited to a maximum of three attempts. Target vulnerable environments are deployed on virtual machines hosted on Amazon Lightsail~\cite{cloud_server}, each configured to emulate a real-world vulnerable application scenario. Generated exploit scripts are automatically executed in the target environments, and server responses are collected for subsequent validation and refinement.

Since the used LLM-based validator is not fully reliable, we perform ground-truth validation to determine exploit correctness. Most cases are validated using rule-based checks (e.g., verifying successful file writes or account creation), while a small number involving complex side effects, such as webshell deployment, require manual inspection. All reported attack success rates are based on ground-truth validation.

To mitigate stochasticity in LLM outputs, each experiment is repeated three times and we report the mean attack success rate (ASR) across runs. Overall, the benchmark includes 104 vulnerabilities and 29 distinct attack objectives, yielding 660 concrete exploitation tasks. We evaluate seven tools across four backbone LLMs, resulting in \textbf{55,440} tool–model–task executions (\textbf{660} tasks × \textbf{7} tools × \textbf{4} models × \textbf{3} runs).

\subsection{RQ1: \tool Performance}\label{sec:rq1}

We evaluate the stage-level and category-level performance of \tool across seven vulnerability information categories using four different LLM backbones. Table~\ref{tab:rq1} reports the results for \textit{Phase A: Trigger Function Construction}, \textit{Phase B: Runtime Exploitation}, and the \textit{Entire Pipeline}. For \textit{Phase~A}, we report the validation pass rate of generated trigger functions after iterative refinement, while for \textit{Phase~B} and entire pipeline, we report the attack success rate (ASR).

Overall, \tool demonstrates strong and stable end-to-end effectiveness, achieving an average ASR of 0.82 across all models and categories. At the stage level, \textit{Phase~A} reaches an average ASR of 0.87, indicating that trigger functions can be constructed reliably from heterogeneous vulnerability information. \textit{Phase~B} further maintains a high average ASR of 0.91, suggesting that once valid trigger logic is available, executable exploits can generally be synthesized and succeed in real environments. Among the evaluated backbones, \textit{DeepSeek-V3} achieves the best overall performance with an entire pipeline ASR of 0.88, followed closely by \textit{Qwen-Plus} at 0.87. \textit{GPT-4o} remains stable but slightly weaker at 0.80, whereas \textit{Claude-3.7} shows a lower pipeline result of 0.72.

A closer inspection reveals that the performance gap of \textit{Claude-3.7} mainly arises in \textit{Phase~A}, where overall performance drops to 0.77. Manual inspection suggests that this degradation is primarily caused by Claude’s stricter safety and content moderation policies, which frequently refuse to generate intermediate attack-related content in \textit{\agent{1} Trigger Logic Extractor}. Once trigger functions are successfully generated, Claude exhibits competitive performance in \textit{Phase~B}, indicating that the observed gap is driven by policy-induced constraints rather than intrinsic reasoning or execution capability limitations. From a category perspective, performance remains balanced across most vulnerability information types. In \textit{Phase~A}, categories containing explicit natural language information (\textit{NL}, \textit{NL+HTTP}, \textit{NL+Payload}) consistently achieve higher ASR than purely structured or code-heavy inputs, reflecting the benefit of semantic redundancy in trigger abstraction. In \textit{Phase~B}, the highest average ASR is observed in the \textit{NL+Payload} category (0.95), indicating that combining semantic context with concrete payload examples provides strong guidance for exploit synthesis. Entire pipeline performance remains stable across categories, with ASR ranging between 0.77 and 0.92, suggesting that \tool generalizes well across heterogeneous vulnerability information.

\begin{table}[!t]
    \centering
    \scriptsize
    \setlength{\tabcolsep}{2.5pt}
	\caption{Performance of \tool across different vulnerability information categories. \textbf{Bold font} highlights the overall best performance across all models. \textcolor{gray}{Grey boxes} highlight the best performance across all categories. \textit{N-TP} denotes \textit{NL+HTTP}, \textit{N-ad} denotes \textit{NL+Payload}, and \textit{N-de} denotes \textit{NL+Code}.}
        \resizebox{\columnwidth}{!}{%
        \begin{tabular}{c|c|ccccccc|c}
	\toprule
	   \multirow{2.5}*{\textbf{Phase}} & \multirow{2.5}*{\textbf{Model}} & \multicolumn{7}{c|}{\textbf{Category}} & \multirow{2.5}*{\textbf{Overall}} \\ 
	   \cmidrule(lr){3-9} & & \textit{NL} & \textit{HTTP} & \textit{Payload} & \textit{Code} & \textit{N-TP} & \textit{N-ad} & \textit{N-de} &  \\
	   \midrule
	   
	   \multirow{5.5}{*}{\shortstack{\textit{Phase A:} \\ \textit{Trigger} \\ \textit{Function} \\ \textit{Construction}}} & \textit{Qwen-Plus} & \textbf{1.00} & \textbf{0.97} & 0.88 & 0.67 & \textbf{1.00} & \textbf{1.00} & \textbf{1.00} & \textbf{0.96} \\
	   & \textit{GPT-4o}  & \textbf{1.00} & 0.84 & 0.65 & \textbf{1.00} & 0.88 & 0.71 & \textbf{1.00} & 0.83 \\
	   & \textit{DeepSeek-V3}  & \textbf{1.00} & 0.92 & 0.94 & \textbf{1.00} & 0.91 & 0.86 & \textbf{1.00} & 0.92 \\
	   & \textit{Claude-3.7}  & \textbf{1.00} & 0.84 & \textbf{1.00} & 0.67 & 0.63 & 0.57 & 0.33 & 0.77 \\
           \cmidrule{2-10}
           & \textit{Avg.}  & \cellcolor{lightgray!40}{1.00} & 0.89 & 0.87 & 0.83 & 0.88 & 0.85 & 0.79 & 0.87 \\
        
        \midrule
        \multirow{5.5}{*}{\shortstack{\textit{Phase B:} \\ \textit{Runtime} \\ \textit{Exploitation}}} & \textit{Qwen-Plus} & 0.94 & 0.87 & 0.88 & 0.93 & 0.92 & 0.85 & 0.80 & 0.89 \\
	   & \textit{GPT-4o}  & 0.91 & \textbf{0.98} & 0.95 & \textbf{1.00} & 0.91 & 0.94 & \textbf{1.00} & \textbf{0.95} \\
	   & \textit{DeepSeek-V3}  & \textbf{0.98} & 0.90 & 0.90 & 0.85 & \textbf{0.94} & \textbf{1.00} & 0.75 & 0.92 \\
	   & \textit{Claude-3.7}  & 0.85 & 0.89 & \textbf{0.98} & 0.87 & 0.82 & \textbf{1.00} & \textbf{1.00} & 0.89 \\
           \cmidrule{2-10}
           & \textit{Avg.}  & 0.92 & 0.91 & 0.93 & 0.91 & 0.90 & \cellcolor{lightgray!40}{0.95} & 0.89 & 0.91 \\

        \midrule
        \multirow{5.5}{*}{\shortstack{\textit{Entire} \\ \textit{Pipeline}}} & \textit{Qwen-Plus} & 0.94 & \textbf{0.85} & 0.75 & 0.70 & \textbf{0.92} & 0.85 & 0.80 & 0.87 \\
	   & \textit{GPT-4o}  & 0.91 & 0.77 & 0.68 & \textbf{1.00} & 0.80 & 0.71 & \textbf{1.00} & 0.80 \\
	   & \textit{DeepSeek-V3}  & \textbf{0.98} & 0.84 & 0.87 & 0.85 & 0.86 & \textbf{0.88} & 0.75 & \textbf{0.88} \\
	   & \textit{Claude-3.7}  & 0.85 & 0.75 & \textbf{0.98} & 0.65 & 0.54 & 0.66 & 0.59 & 0.72 \\
           \cmidrule{2-10}
           & \textit{Avg.}  & \cellcolor{lightgray!40}{0.92} & 0.80 & 0.82 & 0.80 & 0.78 & 0.77 & 0.78 & 0.82 \\
	   \bottomrule
	\end{tabular}}
	\label{tab:rq1}
\end{table}

\subsection{RQ2: Comparative Effectiveness}\label{sec:rq2}

Table~\ref{tab:rq2} summarizes the comparative effectiveness of \tool against all baselines across seven attack objective categories and four backbone LLMs. Overall, \tool consistently outperforms all competing methods under every backbone setting, achieving substantially higher ASR both at the category level and in aggregate. Under the strongest backbone configuration, \textit{DeepSeek-V3}, \tool achieves the highest overall ASR of 0.88, whereas the best-performing baseline, \textit{PentestGPT-Auto}, reaches only 0.34. Similar performance gaps persist across other backbones. Under \textit{Qwen-Plus}, \tool achieves an overall ASR of 0.87 while all baselines remain below 0.27. Under \textit{GPT-4o}, \tool maintains an ASR of 0.80, compared with a maximum baseline performance of 0.32. Even under the most restrictive backbone, \textit{Claude-3.7}, \tool still attains an overall ASR of 0.72, nearly doubling the strongest baseline result of 0.38. Beyond aggregate performance, \tool maintains consistently strong ASR across most objective categories under all backbones, with many categories remaining above 0.75 even when baseline methods rarely exceed 0.35. In particular, \textit{Sensitive Information (SI)} and \textit{SQL Injection (SQLi)} exhibit stable and high performance across all backbones, while \textit{Arbitrary File Write (AFW)} and \textit{Privilege Escalation (PE)} remain strong under \textit{Qwen-Plus}, \textit{GPT-4o}, and \textit{DeepSeek-V3}. Some category-level variations are nevertheless observed. \textit{Webshell (WS)} shows a noticeable drop under \textit{DeepSeek-V3} and \textit{Claude-3.7}, \textit{Remote Command Execution (RCE)} degrades under the more constrained \textit{Claude-3.7} backbone, and \textit{MISC} objectives exhibit slightly lower success rates under \textit{GPT-4o}. These patterns reflect the additional difficulty of synthesizing semantically correct executable payloads while satisfying low-level encoding, serialization, and environment-dependent constraints.

The performance advantage of \tool primarily stems from its explicit extraction of vulnerability trigger logic from heterogeneous and unstructured vulnerability descriptions. By encapsulating vulnerability information into reusable trigger functions, \tool provides a structured and verifiable foundation for downstream exploit instantiation, enabling the system to preserve vulnerability-sensitive constraints and progressively adapt executions using runtime feedback. In contrast, baseline methods directly prompt the LLM to generate complete exploits in a single step, even when augmented with iterative regeneration based on execution feedback. Without explicit modeling of trigger logic or structured intermediate representations, these approaches are more prone to hallucinated parameters, malformed payloads, and environment mismatches, resulting in unstable behavior across attack objectives and backbone settings. Table~\ref{tab:rq2-iterative} further reports the average number of refinement iterations required for successful cases. Overall, \tool achieves higher success rates with fewer refinement iterations, requiring only 2.24 iterations per successful case on average, compared with 2.48 for the most efficient baseline, \textit{PentestAgent-Auto}. Notably, \textit{\tool{}-PhaseA} consistently requires fewer iterations than all baselines, indicating that early-stage trigger function construction effectively constrains downstream exploit instantiation and mitigates unnecessary or hallucinated revisions. Moreover, baseline methods under unconstrained settings exhibit substantially higher iteration overheads (e.g., 13.01 for \textit{PentestGPT} and 35.05 for \textit{VulnBot}), highlighting their instability and inefficiency in practice.

\begin{table}[!t]
    \centering
    \scriptsize
    \setlength{\tabcolsep}{2.5pt}
    \caption{Comparison with baselines across different objective categories. \textbf{Bold font} indicates our \tool's performance. \textcolor{gray}{Grey boxes} highlight the best performance across all baseline methods under the same LLM backbone.}
    
    \resizebox{\columnwidth}{!}{
    \resizebox{\columnwidth}{!}{
    \begin{tabular}{c|c|ccccccc|c}
    \toprule
       \multirow{2.5}*{\textbf{Model}} & \multirow{2.5}*{\textbf{Method}} & \multicolumn{7}{c|}{\textbf{Category}} & \multirow{2.5}*{\textbf{Overall}} \\ 
       \cmidrule(lr){3-9}
       & & \textit{SI} & \textit{RCE} & \textit{AFW} & \textit{PE} & \textit{WS} & \textit{SQLi} & \textit{MISC} &  \\
       \midrule
       
       \multirow{6.5}*{\textit{Qwen-Plus}}  
       &  \textit{DireLLM} & 0.21 & 0.15 & 0.13 & \cellcolor{lightgray!40}{0.20} & 0.13 & 0.19 & 0.13 & 0.16\\
       & \textit{PentestAgent} & 0.29 & 0.16 & 0.30 & \cellcolor{lightgray!40}{0.20} & \cellcolor{lightgray!40}{0.50} & 0.13 & 0.00 & 0.23  \\
       & \textit{PentestGPT-Auto} & 0.35 & 0.19 & 0.26 & \cellcolor{lightgray!40}{0.20} & 0.00 & 0.25 & 0.17 & 0.20\\
       & \textit{PentestGPT-Auto-3} & 0.17 & 0.07 & 0.16 & 0.00 & 0.11 & 0.08 & 0.17 & 0.11\\
       & \textit{VulnBot} & \cellcolor{lightgray!40}{0.44} & \cellcolor{lightgray!40}{0.20} & \cellcolor{lightgray!40}{0.32} & \cellcolor{lightgray!40}{0.20} & 0.14 & \cellcolor{lightgray!40}{0.42} & \cellcolor{lightgray!40}{0.20} & \cellcolor{lightgray!40}{0.27}\\
       & \textit{VulnBot-3} & 0.07 & 0.00 & 0.05 & 0.00 & 0.00 & 0.11 & 0.01 & 0.03\\
       \cmidrule{2-10}
       & \textbf{\textit{\tool{}}} & \textbf{0.75} & \textbf{0.81} & \textbf{0.93} & \textbf{0.88} & \textbf{0.94} & \textbf{0.97} & \textbf{0.92} & \textbf{0.87} \\
       \midrule

       \multirow{6.5}*{\textit{GPT-4o}}  
       &  \textit{DireLLM} & 0.23 & 0.11 & 0.07 & 0.10 & \cellcolor{lightgray!40}{0.50} & 0.19 & 0.00 & 0.15\\
       & \textit{PentestAgent} & 0.46 & \cellcolor{lightgray!40}{0.23} & 0.07 & \cellcolor{lightgray!40}{0.20} & 0.25 & 0.00 & 0.00 & 0.21 \\
       & \textit{PentestGPT-Auto} & 0.38 & 0.11 & 0.25 & 0.00 & 0.00 & 0.25 & 0.00 & 0.20\\
       & \textit{PentestGPT-Auto-3} & 0.13 & 0.07 & 0.10 & 0.00 & 0.00 & 0.00 & 0.00 & 0.08\\
       & \textit{VulnBot} & \cellcolor{lightgray!40}{0.48} & \cellcolor{lightgray!40}{0.23} & \cellcolor{lightgray!40}{0.38} & 0.19 & 0.20 & \cellcolor{lightgray!40}{0.52} & \cellcolor{lightgray!40}{0.22} & \cellcolor{lightgray!40}{0.32}\\
       & \textit{VulnBot-3} & 0.09 & 0.04 & 0.07 & 0.00 & 0.00 & 0.10 & 0.00 & 0.04 \\
       \cmidrule{2-10}
       & \textbf{\textit{\tool{}}} & \textbf{0.80} & \textbf{0.82} & \textbf{0.75} & \textbf{0.94} & \textbf{0.80} & \textbf{0.76} & \textbf{0.67} & \textbf{0.80} \\
       \midrule

       \multirow{6.5}*{\textit{DeepSeek-V3}}  
       &  \textit{DireLLM} & 0.10 & 0.03 & 0.07 & 0.10 & 0.00 & 0.06 & 0.13 & 0.07\\
       & \textit{PentestAgent} & 0.25 & 0.26 & 0.26 & 0.20 & \cellcolor{lightgray!40}{0.50} & 0.00 & \cellcolor{lightgray!40}{0.50} & 0.23 \\
       & \textit{PentestGPT-Auto} & 0.23 & 0.39 & 0.35 & \cellcolor{lightgray!40}{0.40} & 0.34 & 0.38 & 0.32 & 0.34\\
       & \textit{PentestGPT-Auto-3} & 0.23 & 0.09 & 0.15 & 0.10 & 0.16 & 0.13 & 0.13 & 0.14\\
       & \textit{VulnBot} & \cellcolor{lightgray!40}{0.50} & \cellcolor{lightgray!40}{0.41} & \cellcolor{lightgray!40}{0.38} & 0.26 & 0.19 & \cellcolor{lightgray!40}{0.51} & 0.23 & \cellcolor{lightgray!40}{0.35}\\
       & \textit{VulnBot-3} & 0.09 & 0.00 & 0.07 & 0.00 & 0.00 & 0.11 & 0.00 & 0.04\\
       \cmidrule{2-10}
       & \textbf{\textit{\tool{}}} & \textbf{0.90} & \textbf{0.78} & \textbf{0.93} & \textbf{0.94} & \textbf{0.57} & \textbf{0.83} & \textbf{0.83} & \textbf{0.88} \\
       \midrule

       \multirow{6.5}*{\textit{Claude-3.7}}  
       &  \textit{DireLLM} & 0.13 & 0.10 & 0.09 & 0.10 & 0.25 & 0.06 & 0.00 & 0.11\\
       & \textit{PentestAgent} & 0.29 & 0.10 & 0.07 & 0.00 & 0.25 & 0.00 & 0.00 & 0.13 \\
       & \textit{PentestGPT-Auto} & 0.34 & \cellcolor{lightgray!40}{0.45} & 0.28 & \cellcolor{lightgray!40}{0.30} & \cellcolor{lightgray!40}{0.34} & 0.43 & \cellcolor{lightgray!40}{0.50} & \cellcolor{lightgray!40}{0.38}\\
       & \textit{PentestGPT-Auto-3} & 0.17 & 0.05 & 0.22 & 0.20 & 0.17 & 0.07 & 0.00 & 0.12\\
       & \textit{VulnBot} & \cellcolor{lightgray!40}{0.54} & 0.32 & \cellcolor{lightgray!40}{0.43} & 0.23 & 0.23 & \cellcolor{lightgray!40}{0.58} & 0.25 & 0.37\\
       & \textit{VulnBot-3} & 0.10 & 0.05 & 0.08 & 0.00 & 0.00 & 0.11 & 0.00 & 0.05\\
       \cmidrule{2-10}
       & \textbf{\textit{\tool{}}} & \textbf{0.70} & \textbf{0.65} & \textbf{0.80} & \textbf{0.58} & \textbf{0.60} & \textbf{0.76} & \textbf{0.75} & \textbf{0.72} \\
       \bottomrule
    \end{tabular}}
    }
    
    \label{tab:rq2}
\end{table}

\subsection{RQ3: Ablation Study}
\label{sec:ablations}

\begin{table}[!t]
    \centering
    \footnotesize
    \setlength{\tabcolsep}{3pt}   
    \renewcommand{\arraystretch}{0.85}
    \caption{Average number of refinement iterations. \textbf{Bold} values indicate the overall performance of our \tool. \textcolor{gray}{Gray-shaded cells} highlight the least number of iterations among baselines.}
    \resizebox{\columnwidth}{!}{
    \begin{tabular}{c|ccccccc|c}
    \toprule
       \multirow{2.5}{*}{\textbf{Method}} & \multicolumn{7}{c|}{\textbf{Category}} & \multirow{2.5}{*}{\textbf{Overall}} \\ 
       \cmidrule(lr){2-8}
        & \textit{SI} & \textit{RCE} & \textit{AFW} & \textit{PE} & \textit{WS} & \textit{SQLi} & \textit{MISC} &  \\
       \midrule

       \textit{DireLLM} & 2.33 & 3.00 & 3.00 & 2.77 & 2.75 & 2.33 & 3.00 & 2.74\\
       \textit{PentestAgent} & \cellcolor{lightgray!40}{2.00} & \cellcolor{lightgray!40}{2.77} & 2.70 & 3.00 & 2.77 & \cellcolor{lightgray!40}{2.10} & \cellcolor{lightgray!40}{2.00} & \cellcolor{lightgray!40}{2.48}\\
       \textit{PentestGPT-Auto} & 13.67 & 15.44 & 11.20 & 13.67 & 8.50 & 14.83 & 13.75 & 13.01\\
       \textit{PentestGPT-Auto-3} & 2.78 & 2.80 & \cellcolor{lightgray!40}{2.44} & \cellcolor{lightgray!40}{2.60} & \cellcolor{lightgray!40}{2.33} & 2.86 & 2.94 & 2.68\\
       \textit{VulnBot} & 36.83 & 41.56 & 28.15 & 38.81 & 23.04 & 39.92 & 37.02 & 35.05\\
       \textit{VulnBot-3} & 2.99 & 3.00 & 2.64 & 2.80 & 2.51 & 3.00 & 3.00 & 2.88\\
       \cmidrule{1-9}
       \textit{\tool{}-PhaseA} & 1.35 & 1.81 & 1.79 & 2.20 & 1.33 & 1.50 & 1.00 & 1.57\\
       \textit{\tool{}-PhaseB} & 0.43 & 1.04 & 1.00 & 1.40 & 0.33 & 0.50 & 0.00 & 0.67\\
        \textbf{\textit{\tool{}-Entire}} & \textbf{1.78} & \textbf{2.85} & \textbf{2.79} & \textbf{3.60} & \textbf{1.67} & \textbf{2.00} & \textbf{1.00} & \textbf{2.24}\\
       \bottomrule
    \end{tabular}}
    \label{tab:rq2-iterative}
\end{table}

\begin{table}[!t]
    \centering
    \footnotesize
    \renewcommand{\arraystretch}{0.85}
    \caption{Ablation study under different settings.} 
    \begin{tabular}{c|ccc|c}
        \toprule
         {\textbf{LLM}} & \textit{w/o PhaseA} & \textit{w/o \agent{2}\agent{3}} & \textit{w/o \agent{5}\agent{6}}  & \tool\\
         \midrule 
         \textit{Qwen-Plus} & 0.60 & 0.82 & 0.42 & 0.87\\
         \textit{GPT-4o} & 0.55 & 0.74 & 0.14 & 0.80\\
         \textit{DeepSeek-V3} & 0.52 & 0.75 & 0.14 & 0.88\\
         \textit{Claude-3.7} & 0.39 & 0.68 & 0.35 & 0.72\\
         \bottomrule
         
    \end{tabular}
    \label{tab:ablation}
\end{table}

To understand the contribution of each component within the \tool{} framework and validate our design choices, we conduct a systematic ablation study. By selectively disabling or modifying key modules, we quantify their individual impact on overall performance and gain insights into why and how the approach succeeds. All experiments are performed under identical evaluation settings (\S\ref{sec:experiment-setting}), ensuring consistent and comparable results.
All results are shown in Table~\ref{tab:ablation}.

\noindent\textbf{Effect of the \textit{Trigger Function Construction} phase (\textit{w/o PhaseA}).}
This part evaluates the impact of removing the entire \textit{Phase A: Trigger Function Construction}. In this case, the final exploit is generated directly based on the vulnerability information, without any intermediate function abstraction. The results demonstrate that the trigger function construction significantly contributes to the overall performance. Specifically, the ASR under \textit{w/o PhaseA} drops to 0.60 on \textit{Qwen-Plus}, 0.55 on \textit{GPT-4o}, 0.52 on \textit{DeepSeek-V3}, and 0.39 on \textit{Claude-3.7}, indicating that the absence of trigger function leads to a substantial degradation in exploit quality. The impact is particularly notable when using \textit{DeepSeek-V3}, where the ASR decreases from 0.88 to 0.52, highlighting the importance of structured abstraction and refinement in stabilizing downstream exploit generation.

\noindent\textbf{Effect of the \textit{Trigger Function Validator and Refiner} (\textit{w/o \agent{2}\agent{3}}).}
To evaluate the effectiveness of the test-driven validation in \textit{Phase A}, we design a variant of \tool directly using the first generated trigger function from \agent{1} without additional validation and refinement, denoted as \textit{w/o \agent{2}\agent{3}}. The comparison results are presented in Table~\ref{tab:ablation}. We observe that removing the two agents results in a decline in ASR, ranging from 0.05 on \textit{Qwen-Plus} to 0.13 on \textit{DeepSeek-V3}.
These results demonstrate the effectiveness of the test-driven validation in improving both the quality of trigger function and the overall exploit generation performance. However, we also observe that its impact is relatively limited in some cases, as LLMs are able to generate correct trigger functions on the first attempt for certain vulnerabilities, reflecting their strong code generation capability.

\noindent\textbf{Effect of the \textit{Exploit Executor and Refiner} (\textit{w/o \agent{5}\agent{6}}).}
Finally, we examine the contribution of the exploit execution and refinement agents, which adapt exploits based on real execution feedback. In this ablation variant, we remove both modules and treat the first generated exploit as the final result. The results show that this leads to the largest drop in ASR. For example, the performance of \textit{DeepSeek-V3} decreases from 0.88 to 0.14, representing a substantial degradation. Similarly, the ASR of \textit{GPT-4o} drops from 0.80 to 0.14. Although \textit{Qwen-Plus} exhibits the smallest decline, its ASR still decreases from 0.87 to 0.42, amounting to nearly a 50\% reduction. These results indicate that, in most cases, \tool cannot generate a correct exploit in a single attempt even when provided with correct trigger functions. This is because payload generation requires adapting to environment-specific details, which often necessitates iterative refinement based on runtime responses.

\subsection{Case Study}\label{sec:case-study}
We present a real exploit generation case based on CVE-2025-24813~\cite{cve202524813} in the Apache Tomcat server (Appendix~\ref{sec:appendix-case-study}). Tomcat is a popular third-party component deployed in real-world web applications, with over 8.1k stars on GitHub~\cite{tomcat}. We set up a web application runs on a vulnerable Apache Tomcat v9.0.97 instance accessible from the external network. The vulnerability allows an attacker to upload a crafted session file via a partial \texttt{PUT} request and trigger deserialization through a manipulated \texttt{JSESSIONID} cookie, leading to remote code execution. Public references provide configuration snippets and partial HTTP examples but no complete exploit script. To trigger this vulnerability, we need to send a partial \texttt{PUT} request to upload a serialized payload to the session storage path and then send a \texttt{GET} request with a crafted cookie to trigger deserialization. The first issue arises in \textit{Phase A}, where the generated request sequence in trigger function omits the required \texttt{JSESSIONID} cookie. The validator detects this error as a cookie mismatch, and the trigger function is corrected by restoring the missing cookie and revalidated until the test passes. After the triggering logic stabilizes, the main difficulty shifts to exploit instantiation.
The trigger function uses \texttt{CommonsCollections1} gadget as default, while the target application lacks a usable deserialization chain for this gadget, causing the exploit to fail despite well-formed requests. Based on execution feedback, the refiner automatically switches to \texttt{CommonsBeanutils1}, which does not impose a strict JDK version requirement. With this change, the same exploit flow successfully triggers deserialization, and all five attack objectives are achieved within three attempts. Manual inspection confirms that the application runs a high-version JDK8, which explains the failure of \texttt{CommonsCollections1} and validates the correctness of the automated adaptation.

\section{Discussion}

\noindent\textbf{Modularity and Groundtruth Substitution.}
An additional advantage of \tool{} lies in its modular workflow design: each agent in the pipeline (e.g., trigger function extractor, validator, and refiner) is functionally isolated and can be selectively replaced by ground-truth components, such as human annotations or stronger expert agents, when such supervision is available.  This property provides a strong pathway for future improvement: models can be bootstrapped with partial groundtruth to increase accuracy, or hybrid systems can integrate automated steps with curated expert outputs. Such flexibility not only enhances performance but also makes \tool{} adaptable to evolving datasets, stronger LLM backends, or domain-specific expert knowledge.

\noindent\textbf{Threats to Validity.}
Several validity threats should be considered when interpreting our results. From an internal validity perspective, evaluations are conducted on Vulhub environments and vulnerability information obtained from official CVE records, including textual descriptions and referenced materials, which may not fully capture the diversity, configuration complexity, and defensive mechanisms of real-world web deployments. We mitigate this risk by selecting a diverse benchmark of 104 vulnerabilities across multiple categories and repeating each experiment three times to reduce stochastic variance. From an external validity perspective, while our study intentionally focuses on web-application vulnerabilities as defined in the threat model, generalizability may still be influenced by the specific deployment environments and the evolving capabilities and safety policies of LLM backbones. To reduce model- and dataset-specific bias, we evaluate multiple model families and heterogeneous vulnerability representations. Regarding construct validity, we primarily adopt Attack Success Rate (ASR) as the evaluation metric, which emphasizes functional exploit correctness with respect to predefined objectives but does not capture secondary properties such as stealth, persistence, or operational risk. To partially address this limitation, we incorporate multi-objective settings where applicable, enabling assessment beyond single-path exploit success and providing a more realistic approximation of practical penetration testing requirements.
\section{Conclusion}

This paper presents \tool, a novel framework for reliable automated exploit generation. By decomposing the process into trigger function construction and attack-specific instantiation, and validating in each phase systematically, \tool addresses key limitations of existing end-to-end approaches. Extensive evaluations demonstrate that \tool significantly improves attack success rates and adaptability across diverse vulnerabilities and environments, advancing the state of automated penetration testing.

\cleardoublepage
\appendix
\section*{Ethical Considerations}
\textbf{This work investigates automated exploit generation for known vulnerabilities in web applications. While such capabilities can improve the efficiency of defensive security assessment, they also pose potential risks if misused, including lowering the barrier for unauthorized exploitation. Our work is explicitly scoped to authorized penetration testing as defined in the threat model. All experiments are conducted in isolated Vulhub environments using vulnerability information obtained from official CVE records. No experiments target production systems, and no private or sensitive data is involved. From a system design perspective, \tool{} prioritizes structured abstraction, validation, and execution transparency rather than unconstrained exploit synthesis. Intermediate representations make exploit generation auditable and support human oversight, while the modular workflow allows safety mechanisms and access controls to be integrated at critical stages. As with any security automation, responsible deployment ultimately depends on appropriate operational safeguards and compliance with legal and organizational policies.}





\bibliographystyle{IEEEtran}
\bibliography{acmart}

\cleardoublepage
\appendix
\section{Case Study}\label{sec:appendix-case-study}
\begin{figure}[H]
    \centering
    \includegraphics[width=\linewidth]{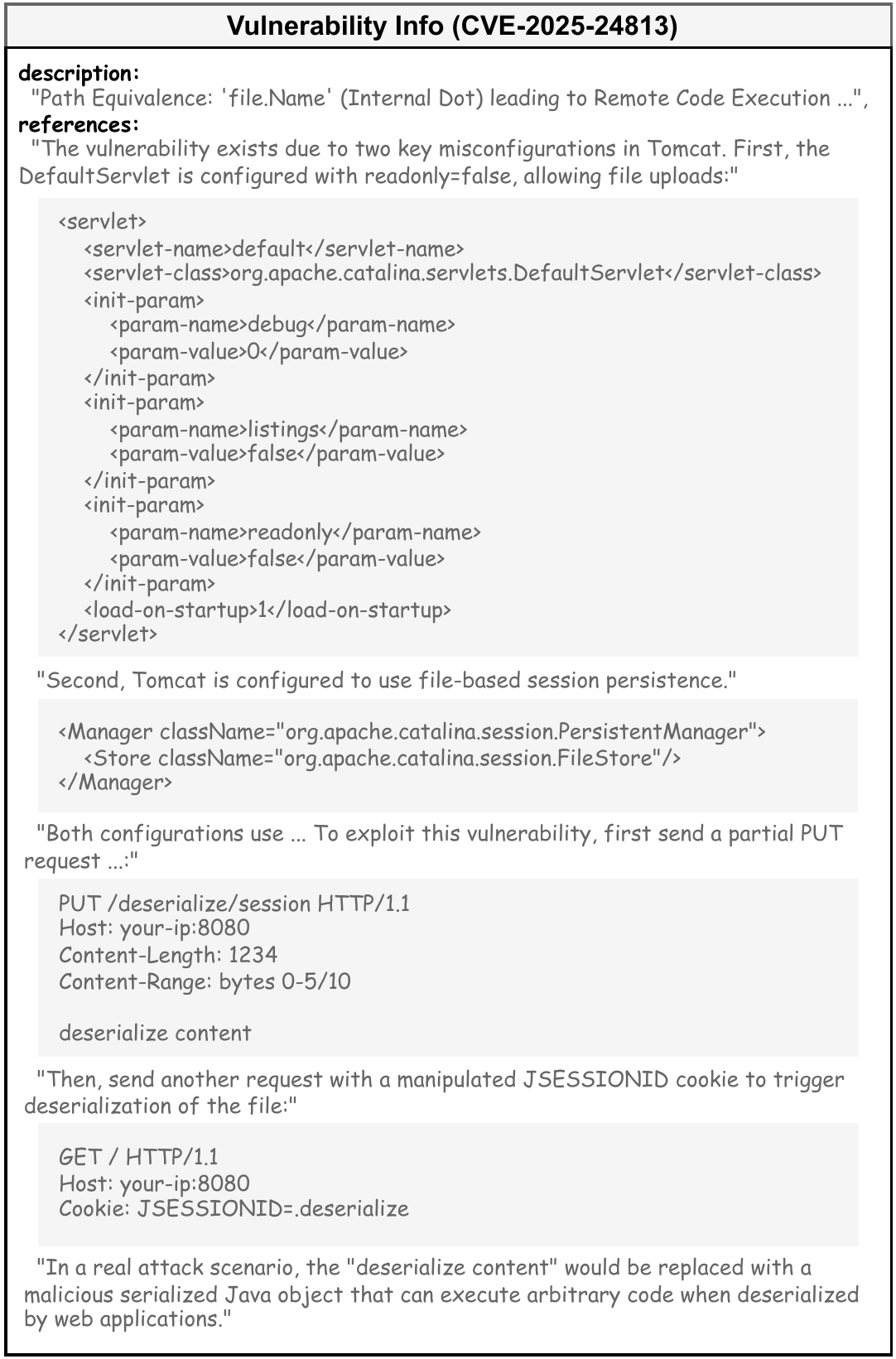}
    \caption{Vulnerability information details for CVE-2025-24813.}
    \label{fig:case-vulinfo}
\end{figure}

\begin{figure}[!t]
    \centering
    \includegraphics[width=\linewidth]{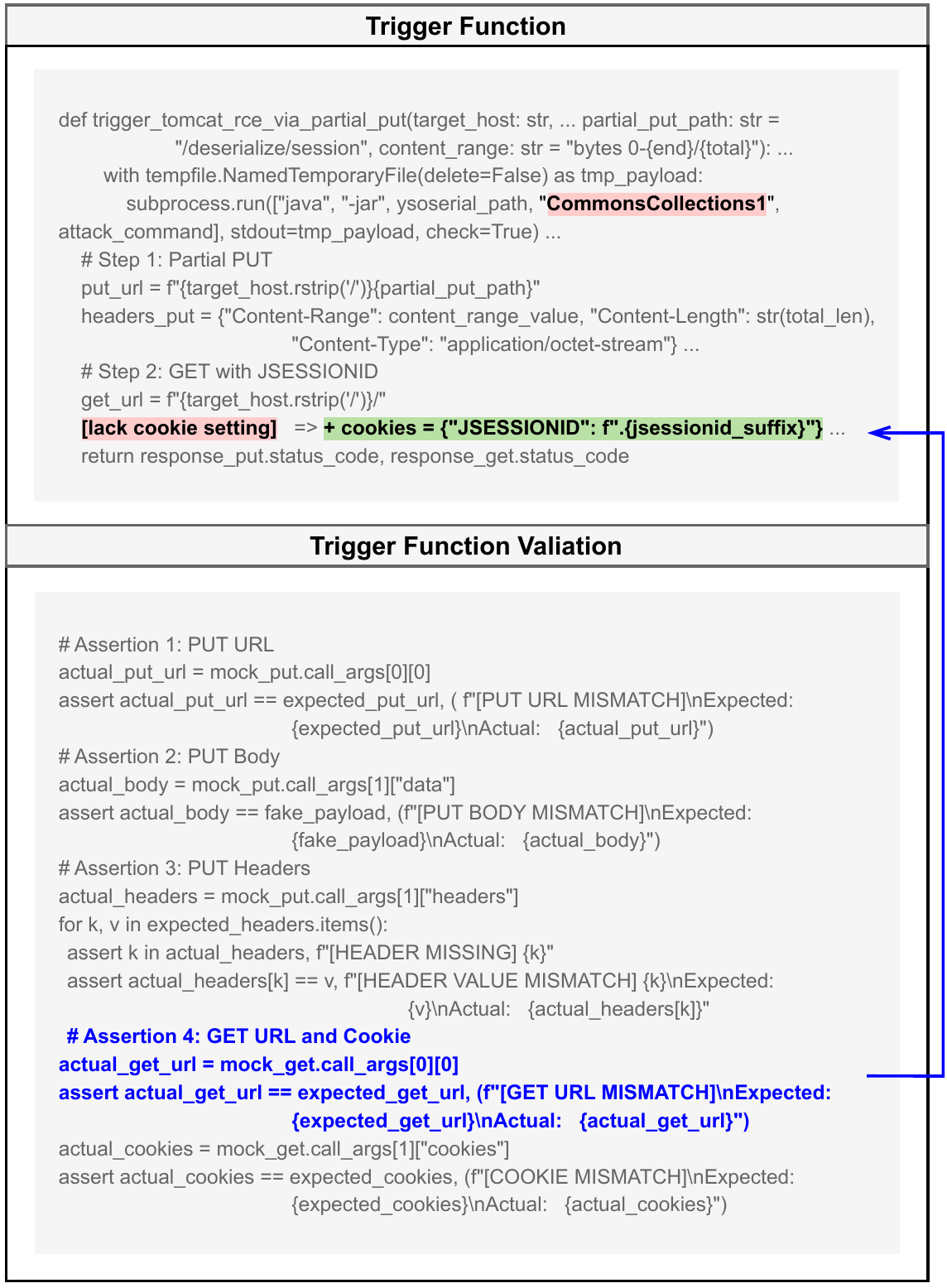}
    \caption{Trigger function construction details for CVE-2025-24813.}
    \label{fig:case-trigger}
\end{figure}

\begin{figure}[!t]
    \centering
    \includegraphics[width=\linewidth]{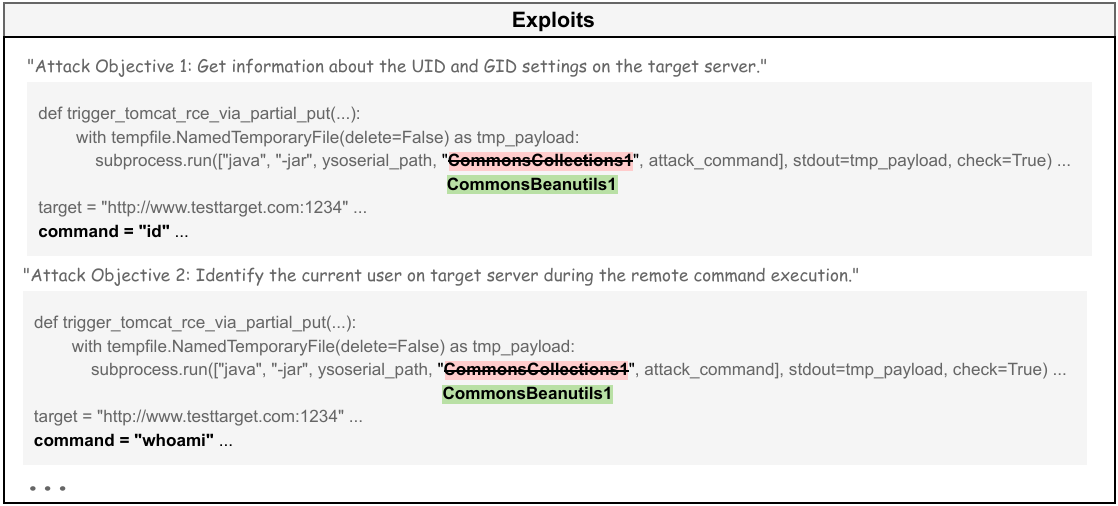}
    \caption{Generated exploits details for CVE-2025-24813.}
    \label{fig:case-exp}
\end{figure}

Figures~\ref{fig:case-vulinfo}, \ref{fig:case-trigger}, and \ref{fig:case-exp} present the code-level details of the case study described in Section~\ref{sec:case-study}.

\section{Examples of Vulnerability Information Representations}
\label{sec:appendix-vuln-examples}

We present representative examples of the four fundamental vulnerability information types used in this work: \textit{natural-language explanations}, \textit{HTTP request examples}, \textit{payload fragments}, and \textit{code snippets}. The remaining three categories (\textit{NL+HTTP}, \textit{NL+Payload}, and \textit{NL+Code}) are mixed cases that combine multiple textual forms, we therefore omit separate examples. All examples are collected from official CVE records and their referenced materials. In all cases, the CVE description alone provides only high-level vulnerability semantics, while concrete triggering logic and exploit construction details appear in the referenced materials.

\begin{figure}[H]
    \centering
    \includegraphics[width=0.85\linewidth]{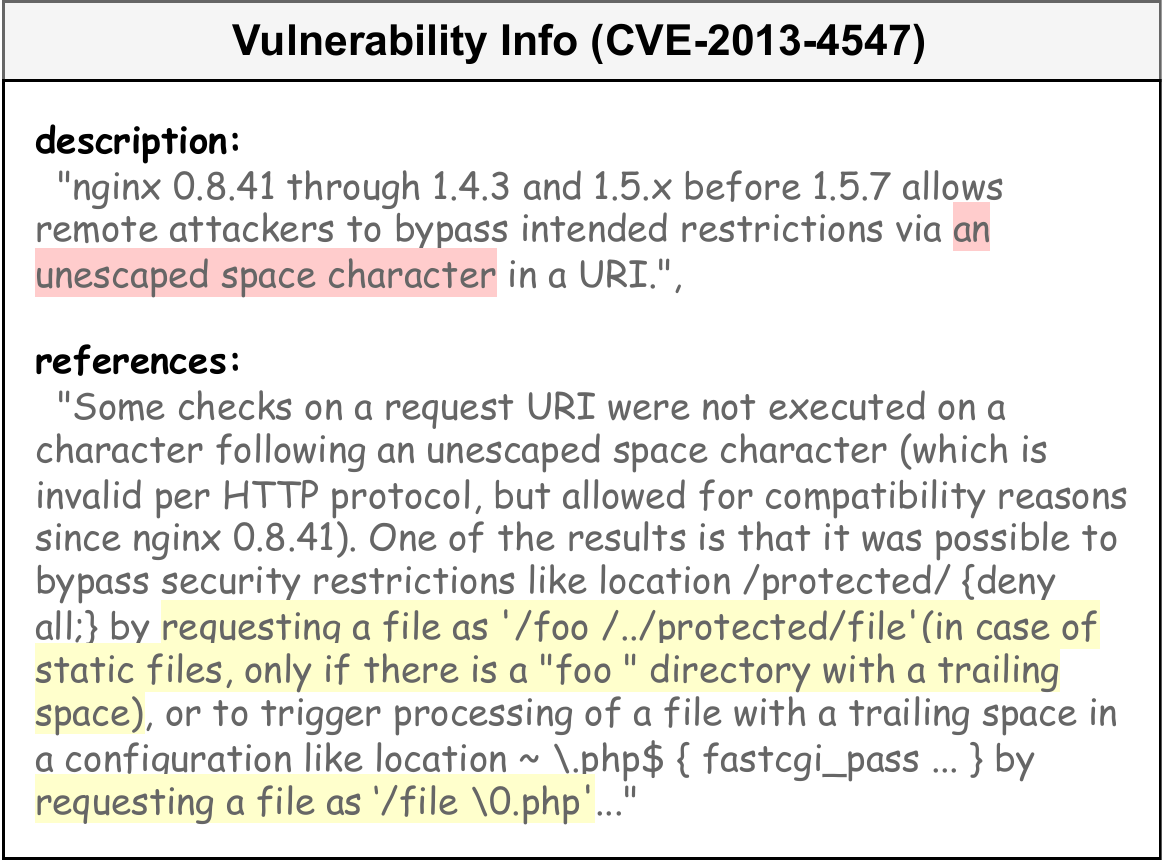}
    \caption{Representative example of natural-language vulnerability information from the CVE-2013-4547 records.}
    \label{fig:appendix-nl}
\end{figure}

\noindent\textbf{Natural-language explanations (CVE-2013-4547).}
Figure~\ref{fig:appendix-nl} shows a case where the CVE description reports that access restrictions can be bypassed via ``an unescaped space character in a URI,'' but does not specify which encodings, path patterns, or request structures realize the bypass. The referenced materials describe the triggering behavior in free-form text and must be interpreted and abstracted before executable logic can be derived.

\begin{figure}[H]
    \centering
    \includegraphics[width=0.85\linewidth]{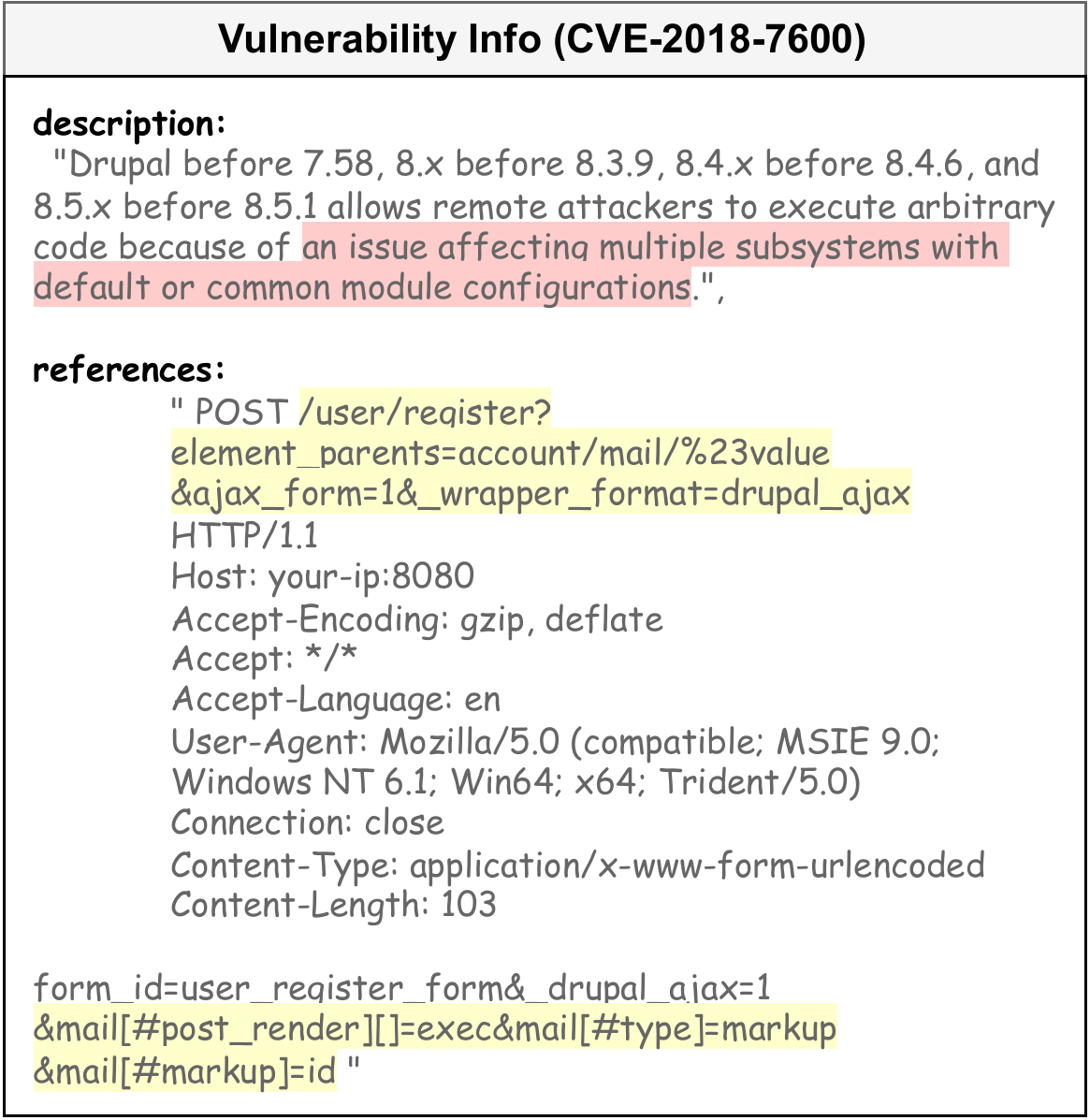}
    \caption{Representative example of HTTP request vulnerability information from the CVE-2018-7600 records.}
    \label{fig:appendix-http}
\end{figure}

\noindent\textbf{HTTP request examples (CVE-2018-7600).}
Figure~\ref{fig:appendix-http} shows a case where the CVE description attributes the vulnerability to ``an issue affecting multiple subsystems with default or common module configurations'' without specifying the concrete misconfigurations or how they enable code execution. The references provide HTTP request instances demonstrating exploitability under specific setups but require abstraction to generalize across environments.

\begin{figure}[H]
    \centering
    \includegraphics[width=0.85\linewidth]{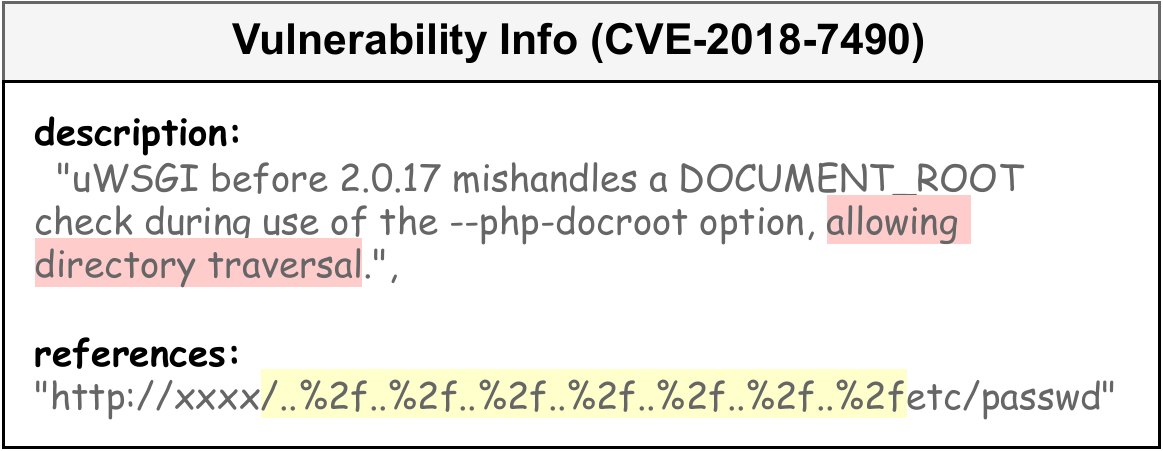}
    \caption{Representative example of payload fragment vulnerability information from the CVE-2018-7490 records.}
    \label{fig:appendix-payload}
\end{figure}

\noindent\textbf{Payload fragments (CVE-2018-7490).}
Figure~\ref{fig:appendix-payload} shows a case where the CVE description reports directory traversal due to improper DOCUMENT\_ROOT checking but does not specify the traversal patterns or payload construction. The references provide partial payload fragments that capture critical exploit components and must be composed and completed before execution.

\begin{figure}[H]
    \centering
    \includegraphics[width=0.85\linewidth]{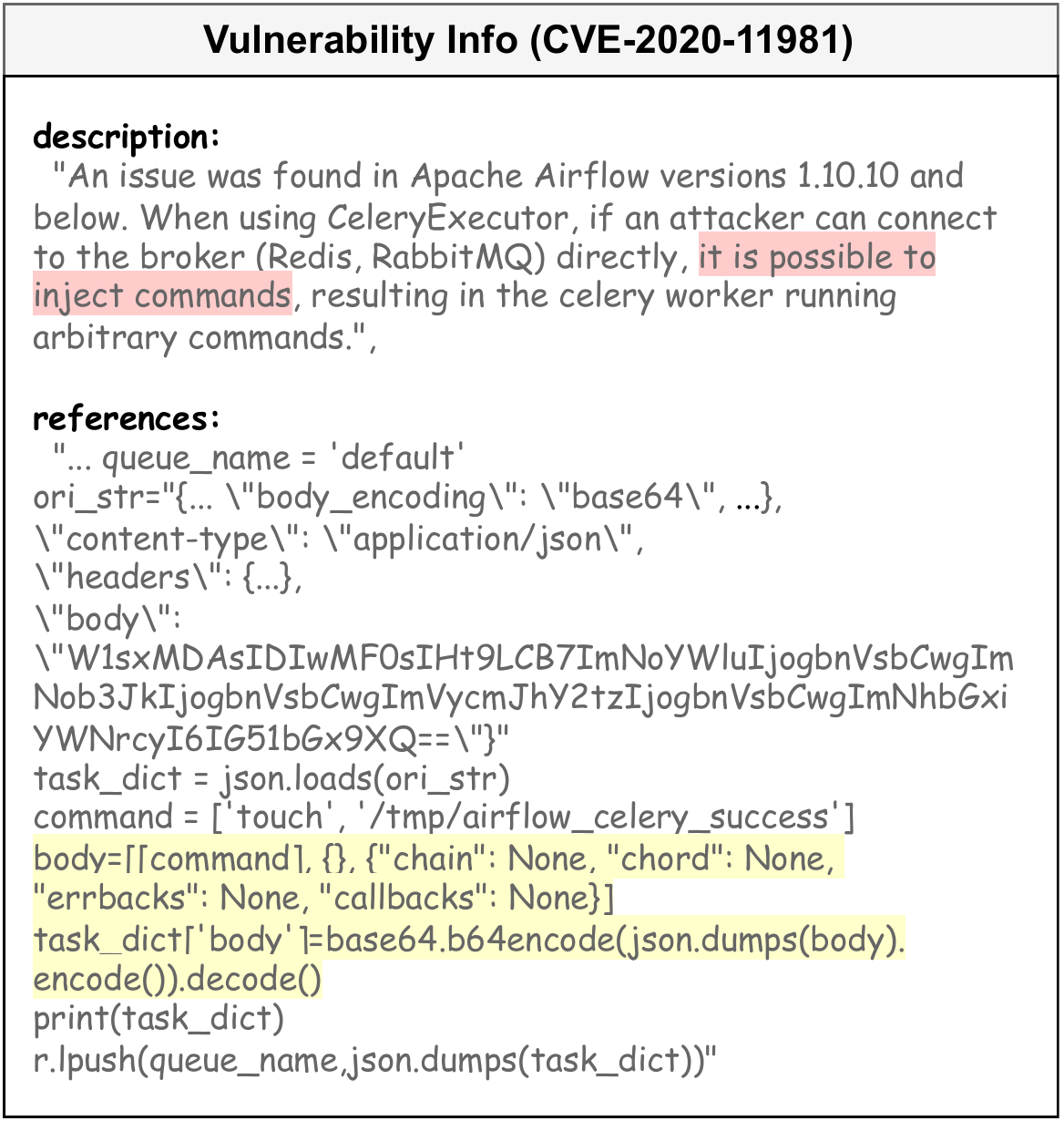}
    \caption{Representative example of code snippet vulnerability information from the CVE-2020-11981 records.}
    \label{fig:appendix-code}
\end{figure}

\noindent\textbf{Code snippets (CVE-2020-11981).}
Figure~\ref{fig:appendix-code} shows a case where the CVE description states that command injection is possible when attackers can connect to the message broker but does not specify how malicious messages are crafted or delivered. The references provide executable code snippets that encode concrete exploitation workflows but require adaptation and validation for reuse in automated pipelines.

\section{Prompt Details for \tool Agents}
\label{sec:appendix-prompts}

This section reports the complete prompts used for LLM interactions within each agent in \tool{} to support reproducibility and detailed inspection. The functional roles and control logic of individual agents, including their non-LLM components (e.g., rule-based validation, exploit execution, and feedback collection), are described in Section~\ref{sec:phase1} and Section~\ref{sec:phase2}. Here we focus exclusively on presenting the exact prompt contents used by each agent. Each figure shows the full system instruction, input specification, and output constraints for the corresponding LLM interaction.

\begin{figure}[H]
    \centering
    \includegraphics[width=0.85\linewidth]{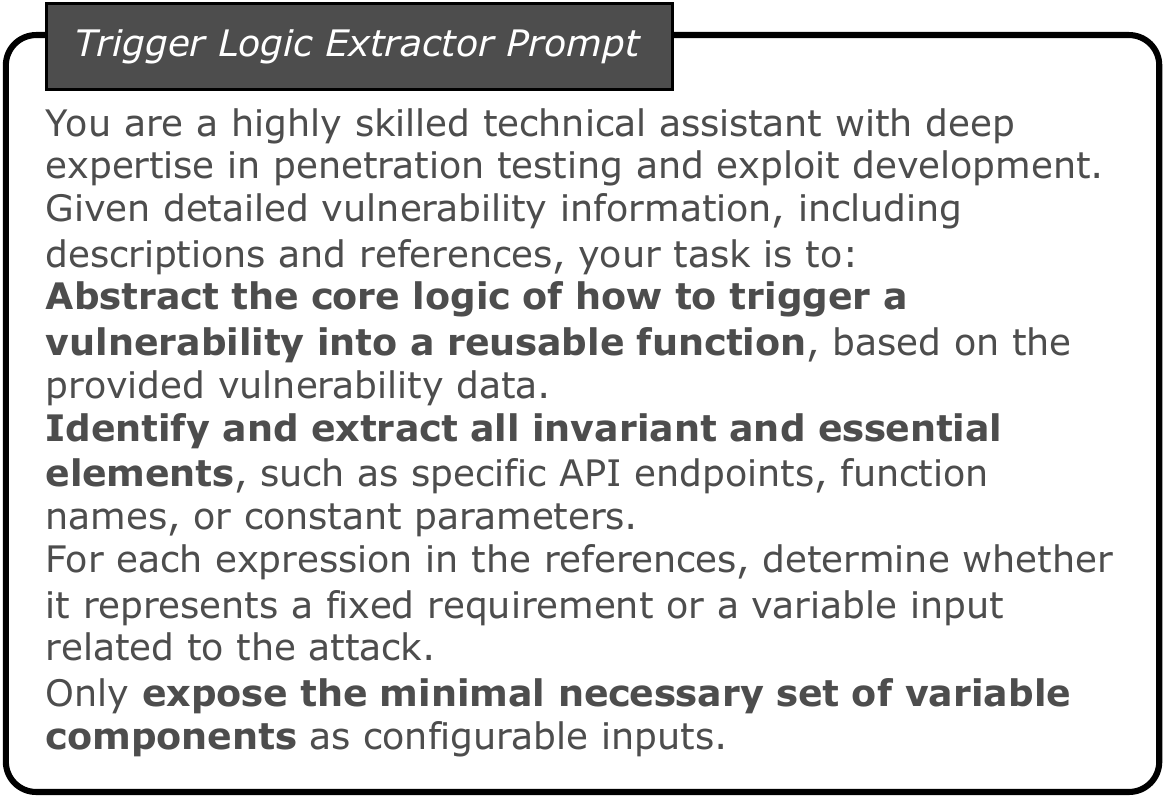}
    \caption{Prompt in the \textit{Trigger Logic Extractor} agent.}
    \label{fig:extractor-prompt2}
\end{figure}

\noindent\textbf{\agent{1} Trigger Logic Extractor.}
Figure~\ref{fig:extractor-prompt2} shows the prompt used by the agent to extract vulnerability trigger logic from the given vulnerability information and to formalize it as an initial trigger function with explicit configurable parameters.

\begin{figure}[H]
    \centering
    \includegraphics[width=0.85\linewidth]{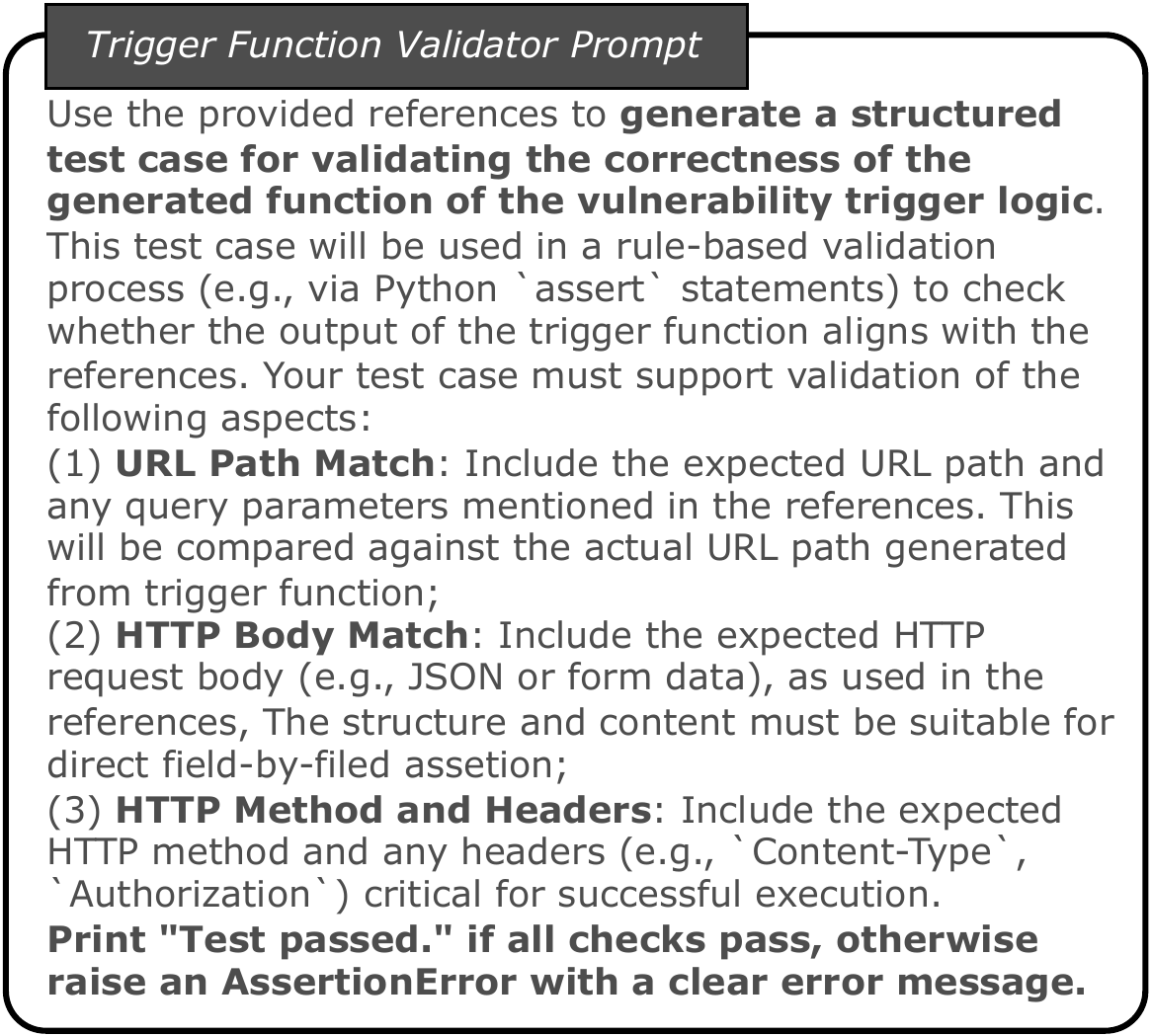}
    \caption{Prompt in the \textit{Trigger Function Validator} agent.}
    \label{fig:funcvalidator-prompt}
\end{figure}

\noindent\textbf{\agent{2} Trigger Function Validator.}
Figure~\ref{fig:funcvalidator-prompt} shows the prompt used by the agent to generate structured test cases from vulnerability references, which are subsequently executed by deterministic assertion checks for trigger function validation.

\begin{figure}[H]
    \centering
    \includegraphics[width=0.85\linewidth]{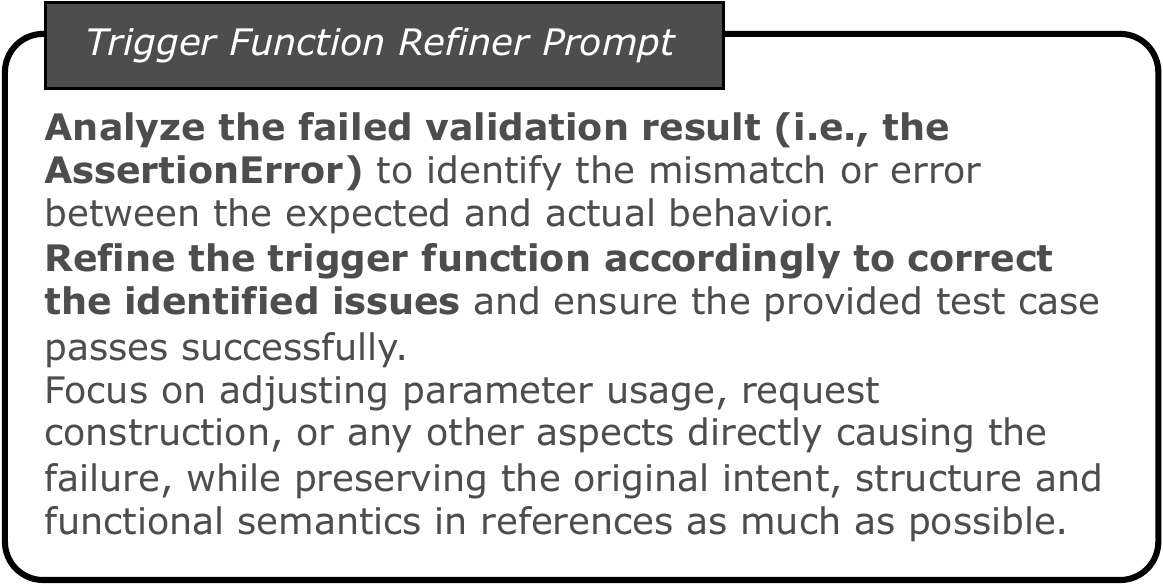}
    \caption{Prompt in the \textit{Trigger Function Refiner} agent.}
    \label{fig:funcrefiner-prompt}
\end{figure}

\noindent\textbf{\agent{3} Trigger Function Refiner.}
Figure~\ref{fig:funcrefiner-prompt} shows the prompt used by the agent to revise a failing trigger function based on assertion feedback, while preserving the original trigger semantics and structure.

\begin{figure}[H]
    \centering
    \includegraphics[width=0.85\linewidth]{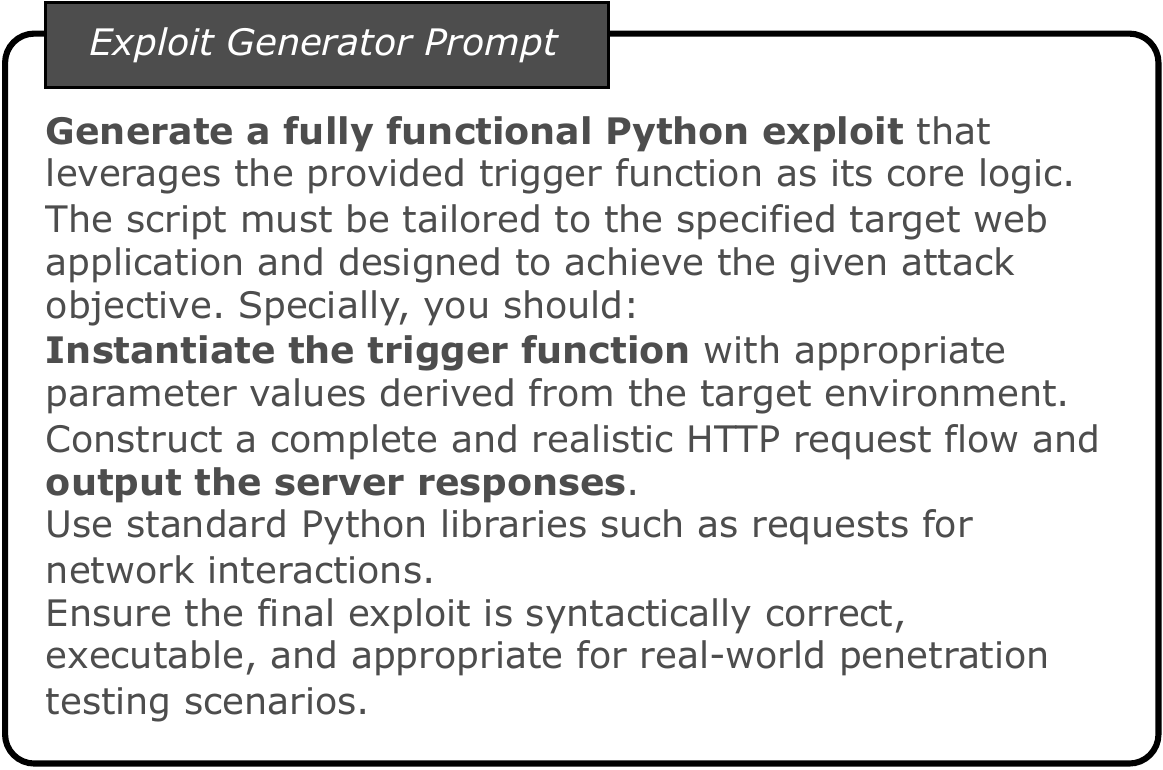}
    \caption{Prompt in the \textit{Exploit Generator} agent.}
    \label{fig:generator-prompt}
\end{figure}

\noindent\textbf{\agent{4} Exploit Generator.}
Figure~\ref{fig:generator-prompt} shows the prompt used by the agent to instantiate a validated trigger function into concrete exploit scripts for specified attack objectives under a given target environment.

\begin{figure}[H]
    \centering
    \includegraphics[width=0.85\linewidth]{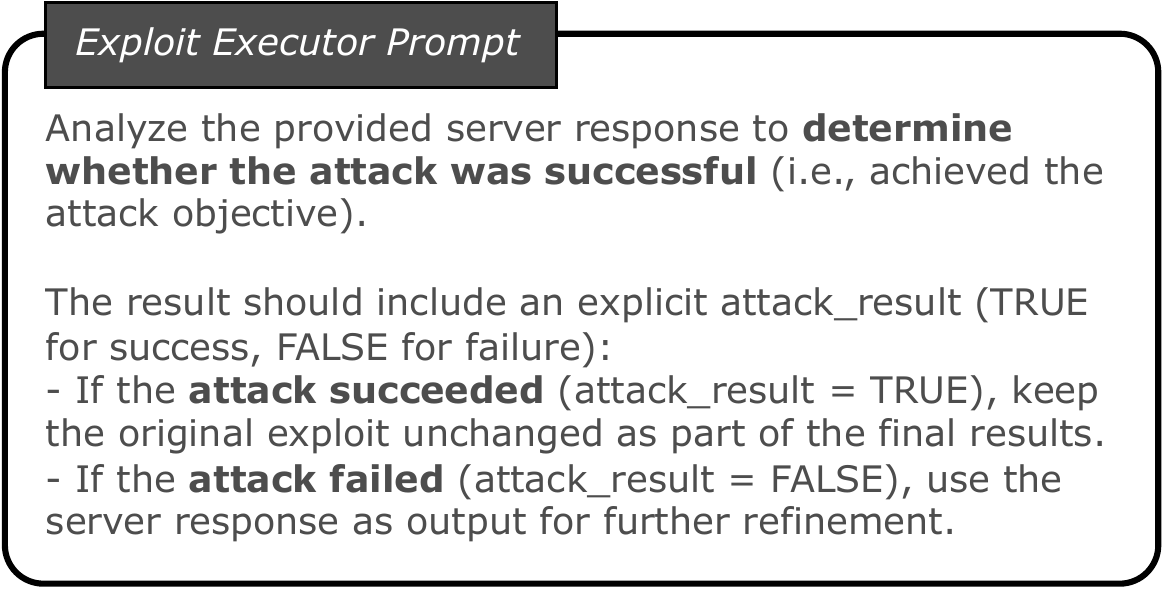}
    \caption{Prompt in the \textit{Exploit Executor} agent.}
    \label{fig:executor-prompt2}
\end{figure}

\noindent\textbf{\agent{5} Exploit Executor.}
Figure~\ref{fig:executor-prompt2} shows the prompt used by the agent to assess runtime execution results with respect to the specified attack objective and produce a binary success decision.

\begin{figure}[H]
    \centering
    \includegraphics[width=0.85\linewidth]{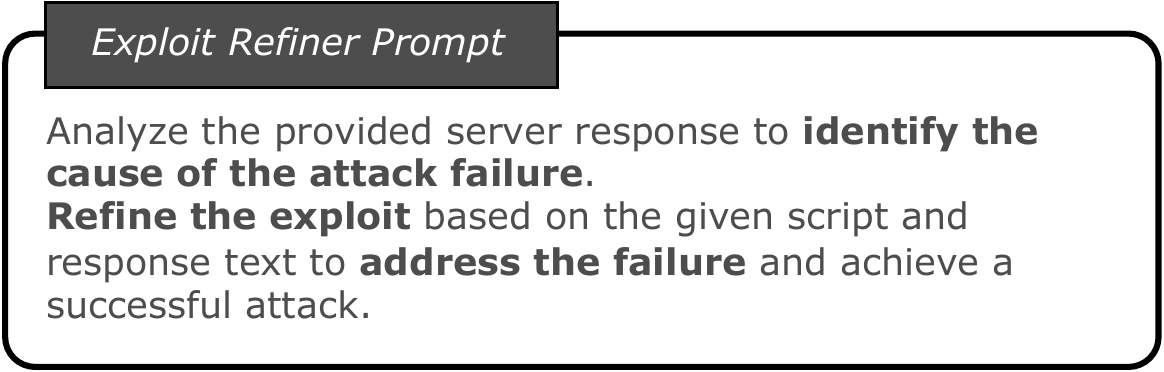}
    \caption{Prompt in the \textit{Exploit Refiner} agent.}
    \label{fig:exprefiner-prompt}
\end{figure}

\noindent\textbf{\agent{6} Exploit Refiner.}
Figure~\ref{fig:exprefiner-prompt} shows the prompt used by the agent to revise failed exploit instances based on execution feedback, while preserving the original attack objective and trigger logic.

\section{Detailed Taxonomy and Dataset Coverage}
\label{sec:appendix-taxonomy}

This section documents the detailed taxonomy and dataset coverage of vulnerability information and attack objectives used in our benchmark. While the construction principles and experimental usage of these taxonomies are described in Section~\ref{sec:experiments}, this section provides the complete categories and CVE mappings to facilitate transparency and reproducibility.

\subsection{Vulnerability Information Categories and CVE Coverage}
\label{sec:appendix-vuln-taxonomy}

Table~\ref{tab:info} reports the complete mapping between vulnerability information categories and their associated CVE identifiers. This table provides an explicit account of dataset coverage and serves as the reference for category-level analysis in RQ1 (Section~\ref{sec:rq1}).

\begin{table}[!t]
    \centering
    \footnotesize
    \caption{Vulnerability information categories and their associated CVE identifiers.}
    \setlength{\tabcolsep}{4pt}
    \renewcommand{\arraystretch}{1.0}
    \begin{tabularx}{\columnwidth}{p{0.28\columnwidth}|X}
        \toprule
        \textbf{Category} & \textbf{Related CVEs} \\
        \midrule

        \textbf{Natural Language (\textit{NL})} &
        CVE-2010-3863, CVE-2013-4547, CVE-2018-3760, CVE-2019-7609, CVE-2020-1957 \\
        \midrule

        \textbf{HTTP Request Examples (\textit{HTTP})} &
        CVE-2012-1823, CVE-2014-3704, CVE-2014-6271, CVE-2017-11610, CVE-2017-12615, CVE-2017-14849, CVE-2017-7525, CVE-2017-8046, CVE-2017-9841, CVE-2018-1000533, CVE-2018-1273, CVE-2018-19475, CVE-2018-19518, CVE-2018-7600, CVE-2018-8715, CVE-2019-10758, CVE-2019-15107, CVE-2019-5418, CVE-2019-7238, CVE-2020-10199, CVE-2020-10204, CVE-2020-13942, CVE-2020-13945, CVE-2020-16846, CVE-2020-17518, CVE-2021-25646, CVE-2021-40438, CVE-2021-40822, CVE-2021-43798, CVE-2022-22963, CVE-2022-24816, CVE-2022-46169, CVE-2023-32315, CVE-2023-4450, CVE-2023-51467, CVE-2024-27348, CVE-2024-4956 \\
        \midrule

        \textbf{Payloads (\textit{Payload})} &
        CVE-2015-3337, CVE-2016-10134, CVE-2017-1000028, CVE-2017-8917, CVE-2018-1000861, CVE-2018-14574, CVE-2018-7490, CVE-2019-14234, CVE-2020-17519, CVE-2021-28164, CVE-2021-28169, CVE-2021-34429, CVE-2021-35042, CVE-2021-41277, CVE-2022-34265, CVE-2023-23752, CVE-2023-25157 \\
        \midrule

        \textbf{Code Snippets (\textit{Code})} &
        CVE-2019-9053, CVE-2020-11981, CVE-2021-26120 \\
        \midrule

        \textbf{\textit{NL+HTTP}} &
        CVE-2014-3120, CVE-2015-1427, CVE-2015-5531, CVE-2016-3088, CVE-2017-10271, CVE-2017-12635, CVE-2017-12636, CVE-2017-15715, CVE-2017-17405, CVE-2019-17558, CVE-2019-3396, CVE-2020-35476, CVE-2020-7012, CVE-2021-26084, CVE-2021-3129, CVE-2021-41773, CVE-2021-42013, CVE-2022-22947, CVE-2022-22965, CVE-2022-26134, CVE-2022-4223, CVE-2023-22515, CVE-2023-22527, CVE-2023-25826, CVE-2023-26360, CVE-2023-28432, CVE-2023-38646, CVE-2023-42793, CVE-2024-36401, CVE-2024-38856, CVE-2024-45507, CVE-2025-24813 \\
        \midrule

        \textbf{\textit{NL+Payload}} &
        CVE-2010-2861, CVE-2015-8562, CVE-2016-1897, CVE-2016-4977, CVE-2020-9402, CVE-2022-22978, CVE-2024-56145 \\
        \midrule

        \textbf{\textit{NL+Code}} &
        CVE-2023-46604, CVE-2024-43441, CVE-2024-45195 \\
        \bottomrule
    \end{tabularx}
    \label{tab:info}
    \vspace{-6pt}
\end{table}

\subsection{Attack Objective Categories and CVE Coverage}
\label{sec:appendix-objective-taxonomy}

Table~\ref{tab:cve} summarizes the attack objective categories, detailed objectives, and the corresponding CVE identifiers included in the benchmark. Each objective is defined in terms of observable runtime behaviors or response conditions, enabling deterministic success evaluation under black-box settings. The table provides a complete mapping between objective categories and vulnerabilities to support consistent evaluation and reproducibility in RQ2 (Section~\ref{sec:rq2}).

\begin{table}[!t]
    \centering
    \footnotesize
    \caption{Average runtime, token consumption, and monetary cost of \tool under different LLM backbones.}
    \label{tab:cost}
    \begin{tabular}{c|ccc}
        \toprule
        \textbf{Model} 
        & \textbf{Avg. Runtime (s)} 
        & \textbf{Avg. Tokens} 
        & \textbf{Avg. Cost (USD)} \\
        \midrule
        
        \textit{Qwen-Plus}     & 610.4  & 38{,}539 & 1.7286 \\
        \textit{GPT-4o}        & 788.7  & 36{,}802 & 0.2257 \\
        \textit{DeepSeek-V3}  & 1{,}729.1 & 55{,}551 & 0.0122 \\
        \textit{Claude-3.7}   & 672.3  & 45{,}244 & 1.3939 \\
        \midrule
        
        \textit{Avg.}          & 950.1  & 44{,}034 & 0.8401 \\
        
        \bottomrule
    \end{tabular}
\end{table}

\begin{table*}[!t]
    \caption{Attack objective categories, detailed objectives, and associated CVE identifiers.}
    \label{tab:cve}
    \centering
    \resizebox{\textwidth}{!}{
    \begin{tabular}{c|l|c}
    \toprule
    \textbf{Categories} & \textbf{Attack Objective Set} & \textbf{Related CVEs}\\
    \midrule
    \makecell{\textbf{Sensitive Information (\textit{SI})}} & \makecell[l]{\textit{- Access the /admin page.}
\\ \textit{- Exfiltrate the content of WEB-INF/web.xml} \\ \textit{- Expose configuration including MySQL username and password.} \\ \textit{- Read the contents of /etc/passwd file from the target server.} \\ \textit{- Retrieve environment variables including `MINIO\_ROOT\_USER`} \\ \textit{and `MINIO\_ROOT\_PASSWORD`.} \\ \textit{- Retrieve the web.xml file.}} & 
\makecell[l]{CVE-2018-3760, CVE-2024-4956, CVE-2021-41277, \\CVE-2017-14849, CVE-2010-2861, CVE-2015-3337, \\CVE-2021-43798, CVE-2019-5418, CVE-2019-3396, \\CVE-2016-1897, CVE-2021-42013, CVE-2018-7490, \\CVE-2015-5531, CVE-2020-17519, CVE-2017-1000028, \\CVE-2023-23752, CVE-2021-28169, CVE-2020-1957, \\CVE-2021-28164, CVE-2021-34429, CVE-2018-8715, \\CVE-2010-3863, CVE-2023-28432, CVE-2022-22978}  \\
    \midrule

    \makecell{\textbf{Remote Command Execution (\textit{RCE})}} & \makecell[l]{\textit{- Get information about the UID and GID settings on the target server.}
\\ \textit{- Identify the current user on target server during the remote command execution.} \\
\textit{- Test external network connectivity with `google.com`.}\\ \textit{- Use evil router to identify current user on target server.} \\ \textit{- Trigger special functions like phpinfo()}} & 
\makecell[l]{CVE-2019-17558, CVE-2023-26360, CVE-2024-38856, \\ CVE-2014-3120, CVE-2022-4223, CVE-2014-6271, \\ CVE-2022-24816, CVE-2022-22947, CVE-2021-26120, \\ CVE-2020-13945, CVE-2021-41773, CVE-2018-7600, \\ CVE-2024-45195, CVE-2022-26134, CVE-2012-1823, \\ CVE-2021-26084, CVE-2022-22965, CVE-2015-1427, \\ CVE-2024-56145, CVE-2019-15107, CVE-2018-19475, \\ CVE-2023-42793, CVE-2023-4450, CVE-2017-12615, \\ CVE-2021-25646, CVE-2024-27348, CVE-2023-22527, \\ CVE-2025-24813, CVE-2016-4977, CVE-2021-3129, \\ CVE-2023-38646, CVE-2020-13942, CVE-2017-17405, \\ CVE-2018-1273, CVE-2024-45507, CVE-2023-46604, \\ CVE-2023-51467, CVE-2020-11981, CVE-2022-46169, \\ CVE-2017-11610, CVE-2017-8046, CVE-2020-35476, \\ CVE-2019-7609, CVE-2018-1000861, CVE-2019-7238, \\ CVE-2020-7012, CVE-2018-1000533, CVE-2019-10758, \\ CVE-2020-10204, CVE-2020-16846, CVE-2020-10199, \\ CVE-2022-22963, CVE-2023-25826, CVE-2017-7525, \\ CVE-2018-19518, CVE-2017-12636, CVE-2013-4547, \\ CVE-2015-8562} \\
    \midrule
    
    \makecell{\textbf{Arbitrary File Write (\textit{AFW})}} & \makecell[l]{\textit{- Write '1234567890' into the file /tmp/test0001 remotely.}
\\ \textit{- Create a new floder /tmp/success remotely.}
\\ \textit{- Create a new file success.txt under the current path remotely.}
\\ \textit{- Execute command `id` and create a new file /tmp/success for the execution result.}
\\ \textit{- Upload a file to /tmp/success on the target server.}} & 
\makecell[l]{CVE-2023-38646, CVE-2020-13942, CVE-2017-17405, \\ CVE-2018-1273, CVE-2024-45507, CVE-2023-46604, \\ CVE-2023-51467, CVE-2020-11981, CVE-2022-46169, \\ CVE-2017-11610, CVE-2017-8046, CVE-2020-35476, \\ CVE-2019-7609, CVE-2018-1000861, CVE-2019-7238, \\ CVE-2020-7012, CVE-2018-1000533, CVE-2019-10758, \\ CVE-2020-10204, CVE-2020-16846, CVE-2020-10199, \\ CVE-2022-22963, CVE-2023-25826, CVE-2017-7525, \\ CVE-2018-19518, CVE-2017-12636, CVE-2020-17518} \\
    \midrule
    
    \makecell{\textbf{Privilege Escalation (\textit{PE})}} & \makecell[l]{\textit{- Bypass the authentication, as evidenced by receiving a 200 status code and } \\ \textit{a session-setting header in the server response.}
\\ \textit{- Create a new account named `testA` with password `test123`.} 
\\ \textit{- Add a system account(testA) and show all accounts on the server from /etc/passwd.}
\\ \textit{- Obtain the password of the admin account.}
\\ \textit{- Use a generated token to access protected endpoints.}} & 
\makecell[l]{CVE-2023-32315, CVE-2024-43441, CVE-2023-22515, \\ CVE-2019-9053, CVE-2017-12635, CVE-2010-3863, \\ CVE-2020-1957, CVE-2022-22978, CVE-2018-8715}  \\
    \midrule
    
    \makecell{\textbf{Webshell (\textit{WS})}} & \makecell[l]{\textit{- Upload a webshell and obtain a connection to target server.}
\\ \textit{- Obtain the reverse shell connection.} } & 
\makecell[l]{CVE-2017-10271, CVE-2017-9841, CVE-2016-3088, \\ CVE-2017-15715, CVE-2020-17518}  \\
    \midrule
    
    \makecell{\textbf{SQL Injection (\textit{SQLi})}} & \makecell[l]{\textit{- Get SQL error message, which confirming a successful SQL injection.}
\\ \textit{- Retrieve the version of PostgreSQL via SQL injection.} \\
\textit{- Obtain table or schema data via SQL injection.}} & 
\makecell[l]{CVE-2020-9402, CVE-2023-25157, CVE-2021-35042, \\ CVE-2016-10134, CVE-2019-14234, CVE-2017-8917, \\ CVE-2013-4547, CVE-2014-3704}  \\
    \midrule
    
    \makecell{\textbf{Miscellaneous (\textit{MISC})}} & \makecell[l]{\textit{- Get response from `google.com`.}
\\ \textit{- Redirect user to `www.google.com`.} \\
\textit{- Trigger the XSS with `hello world` message.}} & 
\makecell[l]{CVE-2021-40822, CVE-2021-40438, CVE-2024-36401, \\ CVE-2018-14574}  \\
\bottomrule
\end{tabular}
}
\end{table*}

\section{Cost Analysis of \tool}
\label{sec:cost}

This section reports the computational and monetary cost of running \tool under different LLM backbones in terms of average wall-clock runtime, token consumption, and monetary cost.\footnote{We use the public pricing at the time of experimentation: \textit{gpt-4o-2024-08-06} (Input: \$1.25 / 1M tokens, Output: \$5.00 / 1M tokens), \textit{deepseek-chat} (Input: ¥0.20 / 1M tokens, Output: ¥3.00 / 1M tokens), \textit{qwen-plus-2025-04-28} (Input: \$0.40 / 1M tokens, Output: \$1.20 / 1M tokens), and \textit{claude-3-7-sonnet-20250219} (Input: \$3.00 / 1M tokens, Output: \$15.00 / 1M tokens). All costs are converted to USD for reporting consistency.} 
All statistics are computed over all executions, including both successful and failed runs. Each execution corresponds to one complete exploit generation process for a single vulnerability, which may internally involve multiple attack objectives and iterative refinements. Table~\ref{tab:cost} summarizes the results.

Overall, \tool exhibits moderate runtime overheads across all models, with average runtimes ranging from several minutes to under half an hour per vulnerability. Runtime differences mainly reflect model inference latency and the number of refinement iterations triggered during execution. \textit{DeepSeek-V3} incurs the longest average runtime and highest token usage, whereas \textit{Qwen-Plus}, \textit{GPT-4o}, and \textit{Claude-3.7} show comparable wall-clock performance. Monetary cost varies substantially due to pricing differences rather than token volume alone: \textit{DeepSeek-V3} achieves the lowest cost despite higher token usage, while \textit{Claude-3.7} and \textit{Qwen-Plus} incur higher costs under premium pricing; \textit{GPT-4o} provides a balanced trade-off between runtime efficiency and cost. These results indicate that \tool can be deployed under different cost--performance constraints while maintaining stable end-to-end behavior.
\end{document}